\documentclass[a4paper,11pt]{article}

\usepackage[normalem]{ulem}

\usepackage{xcolor}
\usepackage[english]{babel}
\usepackage[T1]{fontenc}
\usepackage[utf8]{inputenc}
\usepackage{authblk}
\usepackage{mathtools}
\usepackage{epsfig}
\usepackage{slashed}
\usepackage{amsmath,amssymb}
\usepackage{mathrsfs}
\usepackage{amsfonts}
\usepackage{enumitem}
\usepackage{graphicx,color,xcolor}
\usepackage{cite}
\usepackage{float}
\usepackage{subcaption}
\usepackage{soul}
\usepackage{hyperref}
\usepackage{wrapfig}

\hypersetup{
	colorlinks=true,
	linkcolor=blue,
	filecolor=magenta,
	urlcolor=cyan,
}
\usepackage{cleveref}
\usepackage[left=2.5cm,right=2.5cm,top=2.5cm,bottom=2.5cm]{geometry}
\numberwithin{equation}{section}

\definecolor{refcol}{rgb}{0.9,0.1,0.1}
\hypersetup{colorlinks=true,linkcolor=blue,citecolor=refcol,urlcolor=cyan,linktocpage}

\begin{document}
	\begin{titlepage}
		\thispagestyle{empty}

		\title{
			{\Huge\bf  Freudenthal Duality in Conformal Field Theory}
		}
		
		\vfill
		
		\author{
			{\bf Arghya Chattopadhyay$^a$}\thanks{{\tt \href{mailto:arghya.chattopadhyay@umons.ac.be}{arghya.chattopadhyay@umons.ac.be}}},
			{\bf Taniya Mandal$^b$}\thanks{{\tt
					\href{taniya.mandal@niser.ac.in}{taniya.mandal@niser.ac.in}}},
			{\bf Alessio Marrani$^{c}$}\thanks{{\tt \href{alessio.marrani@um.es}{alessio.marrani@um.es}}}
			\smallskip\hfill\\      	
			\small{
				${}^a${\it Service de Physique de l'Univers, Champs et Gravitation}\\
				{\it Universit\'e de Mons, 20 Place du Parc, 7000 Mons, Belgium}\\
				\hfill\\
				${}^b${\it School of Physical Sciences, National Institute of Science Education and Research}\\
				{\it An OCC of Homi Bhabha National Institute}\\
				{\it Bhubaneswar 752050, India}\\
				\hfill\\
				${}^c${\it Instituto de F\'{\i}sica Teorica, Dep.to de F\'{\i}sica}\\
				{\it Universidad de Murcia, Campus de Espinardo, E-30100, Spain}\\
			}
		}

		\vfill
		
		\date{
			\begin{quote}
				\centerline{{\bf Abstract}}
				{\small
					Rotational Freudenthal duality (RFD) relates two extremal Kerr-Newman (KN) black holes (BHs) with different angular momenta and electric-magnetic charges, but with the same Bekenstein-Hawking entropy. Through the Kerr/CFT correspondence (and its KN extension), a four-dimensional, asymptotically flat extremal KN BH is endowed with a dual thermal, two-dimensional conformal field theory (CFT) such that the Cardy entropy of the CFT is the same as the Bekenstein-Hawking entropy of the KN BH itself. Using this connection, we study the effect of the RFD on the thermal CFT dual to the KN extremal (or doubly-extremal) BH. We find that the RFD maps two different thermal, two-dimensional CFTs with different temperatures and central charges, but with the same asymptotic density of states, thereby matching the Cardy entropy. We also discuss the action of the RFD on doubly-extremal rotating BHs, finding a spurious branch in the non-rotating limit, and determining that for this class of BH solutions the image of the RFD necessarily over-rotates.}
		\end{quote}}
	\end{titlepage}
	\thispagestyle{empty}\maketitle\vfill \eject
	
	\baselineskip=18pt

	\tableofcontents
	

	\section{Introduction}
	
	Freudenthal duality (FD), as originally introduced, is a non-linear,
	anti-involutive map between the electric-magnetic (e.m.) charge
	configurations supporting two static, spherically symmetric extremal black
	holes (BHs) as solutions in four-dimensional $\mathcal{N}=2$ supergravity in
	asymptotically flat space, and defined as the gradient of the entropy in the
	e.m. charge symplectic space \cite%
	{Borsten:2009zy,Ferrara:2011gv,Borsten:2012pd}. The study of this duality
	was further extended for BHs in gauged supergravity \cite{Klemm:2017xxk} and
	in many other contexts \cite%
	{Marrani:2012uu,Galli:2012ji,Fernandez-Melgarejo:2013ksa,Mandal:2017ioi,Borsten:2018djw,Borsten:2019xas}%
	. A remarkable feature of FD is that the BHs have the same
	Bekenstein-Hawking entropy, despite being supported by different,
	Freudenthal-dual, e.m. charges; moreover, they also share the same attractor
	values of the moduli fields at the event horizon. This can be traced back to
	the fact that the Bekenstein-Hawking entropy of the aforementioned class of
	BHs is a homogeneous, degree-two function of its e.m. charges. Hence,
	extending this duality to more generic cases, such as non-extremal and/or
	rotating (stationary) BHs is not straightforward, as their entropy loses the
	above-mentioned homogeneity \cite{Chattopadhyay:2021vog}. Recently, the
	present authors have extended this idea to near-extremal BHs, stating that
	two different near-extremal BHs can share the same entropy when their
	charges are related by a suitable non-linear relation, which necessarily
	yields that they have different temperatures \cite{Chattopadhyay:2022ycb}.
	This (non-anti-involutive) duality has been named the generalised FD (GFD).
	Later on, a further extension to over- and under- rotating stationary
	extremal BHs has been formulated in \cite{Chattopadhyay:2023nfc},
	necessarily yielding different angular momenta for the Freudenthal-dual BHs.
	This (non-anti-involutive) duality has been named rotational FD (RFD). It is interesting to note that entropy stays invariant also for a class of Kerr-Newman-deSitter BHs with the same charge and angular momentum but with different masses \cite{McInerney:2015xwa,Chakrabhavi:2023avi}.
	
	So far, the understanding and the application of GFD/RFD has been limited to
	the macroscopic regime of BHs in Maxwell-Einstein-scalar theories of gravity
	(which are not necessarily supersymmetric, but could be considered as
	bosonic sectors of four-dimensional supergravity\footnote{%
		For the interplay between BPS supersymmetry-preserving properties of
		(extremal) BHs and FD, see Sec. 8.4 of \cite{Borsten:2019xas}}). As a BH
	behaves as a thermal object with a finite temperature and entropy, one might
	wonder to count the degeneracy of its microstates. It is indeed possible to
	count the microstates for supersymmetric, extremal BHs with large charges
	\cite{Strominger:1996sh,Juan-Maldacena-1997,Shih:2005qf}. In this paper, we
	attempt to understand the effect of FD on the microscopic entropy of an
	extremal Kerr-Newman BH in asymptotically flat space in four dimensions.
	Through the Kerr/CFT correspondence \cite{Guica:2008mu}, there exists a
	two-dimensional, thermal conformal field theory (CFT) that is dual to the
	near horizon geometry of the extremal Kerr BH in asymptotically flat space.
	The Cardy entropy, corresponding to the density of states of such a CFT,
	exactly produces the Bekenstein-Hawking entropy of the Kerr BH.
	After the seminal paper \cite{Guica:2008mu}, the Kerr/CFT correspondence has
	been generalized to the presence of Maxwell fields \cite{Haco:2019ggi} and
	thus to the Kerr-Newman extremal BH solutions (which can be regarded as
	solution to the bosonic sector of $\mathcal{N}=2$, $D=4$ \textquotedblleft
	pure\textquotedblright\ supergravity). A consistent formulation within
	(ungauged and gauged) supergravity was already put forward in \cite%
	{Chow:2008dp}, within the so-called extremal black hole/CFT correspondence
	(see also \cite{Compere:2009dp}, and further \cite{Compere:2012jk} for a
	comprehensive review). Here, we will analyze the effect of RFD on
	the density of states of a two-dimensional CFT that is dual to an
	asymptotically flat, generally dyonic Kerr-Newman extremal BH, regarded as
	a solution to a Maxwell-Einstein-scalar system, which can (but does not
	necessarily have to) be conceived as the purely bosonic sector of a ($\mathcal{N}\geqslant 2$-extended, ungauged) supergravity theory in four
	space-time dimensions.
	
	Our plan for the paper is as follows.
	
	We begin with a brief discussion of RFD in \cref{sec:introrfd}, specifically review RFD for extremal black hole in \cref{extremal RFD}. In 
	\cref{app:anotherrfd} we give a general proof of the uniqueness of RFD (as
	well as a treatment of the \textquotedblleft spurious\textquotedblright\
	branch of solutions in the non-rotating limit) for doubly-extremal
	Kerr-Newman BHs, thus completing the analysis given in \cite%
	{Chattopadhyay:2023nfc}; interestingly, we find that any RFD-transformed
	doubly-extremal KN BH is over-rotating.
	After that, in \cref{sec:KNBH} we elaborate the details of the Kerr/CFT
	correspondence (and its aforementioned extensions and generalizations to
	Maxwell-Einstein-scalar systems), focusing our attention specially on the
	duality in presence of (running) scalar fields. Then in \cref{sec:cardy} we discuss the action of the RFD on the Cardy entropy in the CFT dual, both for the extremal and doubly-extremal rotating BHs,
	providing a novel microscopic picture of RFD. We finish with a outlook and
	implications of our findings in \cref{sec:conclusion}.  Finally, for pedagogic completeness,
	in \cref{app:kerrcardy} we provide a concise review of the Cardy
	formula for CFT.
	
	\section{Rotational Freudenthal Duality (RFD)}
	\label{sec:introrfd} 
	In this section, we briefly discuss the effect of Rotational Freudenthal duality on extremal black holes \cite{Chattopadhyay:2023nfc} and also analyze in detail the same for double-extremal rotating black holes.
	\subsection{RFD for Extremal Black Holes}\label{extremal RFD} 
	The Bekenstein-Hawking entropy of an
	asymptotically flat, extremal, under-/over- rotating stationary
	(Kerr-Newman, generally dyonic) charged BH (regarded as a solution to
	Einstein gravity, coupled to Maxwell fields and running scalar fields,
	within the so-called `Maxwell-Einstein-scalar system', which can further be regarded
	as the purely bosonic sector of four-dimensional supergravity theories),
	with e.m. charges collected in the symplectic vector $Q$ and angular
	momentum $J$, is respectively given by \cite{Astefanesei:2006dd,
		Ferrara:2008hwa},
	\begin{eqnarray}
		S_{\text{under}}\left( Q,J\right) &=&\sqrt{S_{0}^{2}\left( Q\right) -J^{2}}
		\label{enunderover} \\
		S_{\text{over}}\left( Q,J\right) &=&\sqrt{J^{2}-S_{0}^{2}\left( Q\right) }
		\label{enunderover2}
	\end{eqnarray}%
	where $S_{0}(Q)$ is the extremal entropy in the absence of rotation. Though
	extremal, the entropies of such extremal BHs lose the homogeneity property
	on $Q$ due to the presence of angular momentum. Hence, a canonical method of
	establishing FD does not hold for them. As mentioned, in \cite%
	{Chattopadhyay:2023nfc} the present authors have formulated a generalised
	version of FD, named RFD, for this class of BHs : two rotating BHs with
	angular momenta $J$ and $J+\delta J$ have the same entropy i.e.
	\begin{equation}
		S_{\text{under(over)}}\left( Q,J\right) =S_{\text{under(over)}}\left( \Omega
		\frac{\partial S_{\text{under(over)}}\left( Q,J+\delta J\right) }{\partial Q}%
		,J+\delta J\right) ,  \label{enequal}
	\end{equation}%
	when their e.m. charges are RFD-dual, ie. they are related by
	\begin{equation}
		Q\rightarrow \hat{Q}(Q,J+\delta J)=\Omega \frac{\partial S_{\text{under(over)%
			}}\left( Q,J+\delta J\right) }{\partial Q}.
	\end{equation}%
	Consistently, the $J\rightarrow 0^{(+)}$ limit of the extremal,
	under-rotating BH entropy is\footnote{%
		When the $U$-duality group is a `generalized group of type $E_{7}$' of
		non-degenerate resp. degenerate type, $S_{0}$ equals (in units of $\pi $) $%
		\sqrt{\left\vert I_{4}(Q)\right\vert }$ or $\left\vert I_{2}(Q)\right\vert $%
		, where $I_{4}(Q)$ and $I_{2}(Q)$ are the (unique, primitive) polynomial of
		e.m. charges (homogeneous of degree four resp. two), invariant under the
		(non-transitive, symplectic) action of the $U$-duality group itself.} $%
	S_{0}\left( Q\right) $. By analysing the solutions to a sextic equation of $\delta J$, it has been
	shown both analytically and numerically that this mapping is unique.
	
	Introducing a real and positive parameter $\alpha$ for the mother black hole as $\alpha=\frac{J^2}{S_0^2}$, one can find that the angular momenta $\hat{J}=J+\delta J$ of the RFD dual black hole behaves as follows \cite{Chattopadhyay:2023nfc}
	\begin{eqnarray}
		\hat{J}^{2} &=&\left( J+\delta J\right) ^{2}=f(\alpha )S_{0}^{2},
		\label{thiss} \\
		f(\alpha ) &:&={\frac{1}{2^{2/3}}}\left( 1+\sqrt{1+{\frac{4}{27}}(\alpha
			-2)^{3}}\right) ^{2/3}+{\frac{1}{2^{2/3}}}\left( 1-\sqrt{1+{\frac{4}{27}}%
			(\alpha -2)^{3}}\right) ^{2/3}+{\frac{1}{3}}(\alpha +1).\notag\\ \label{eq:falpha}
	\end{eqnarray}%
	\begin{figure}[h]
		\centering
		\includegraphics[width=0.8\textwidth]{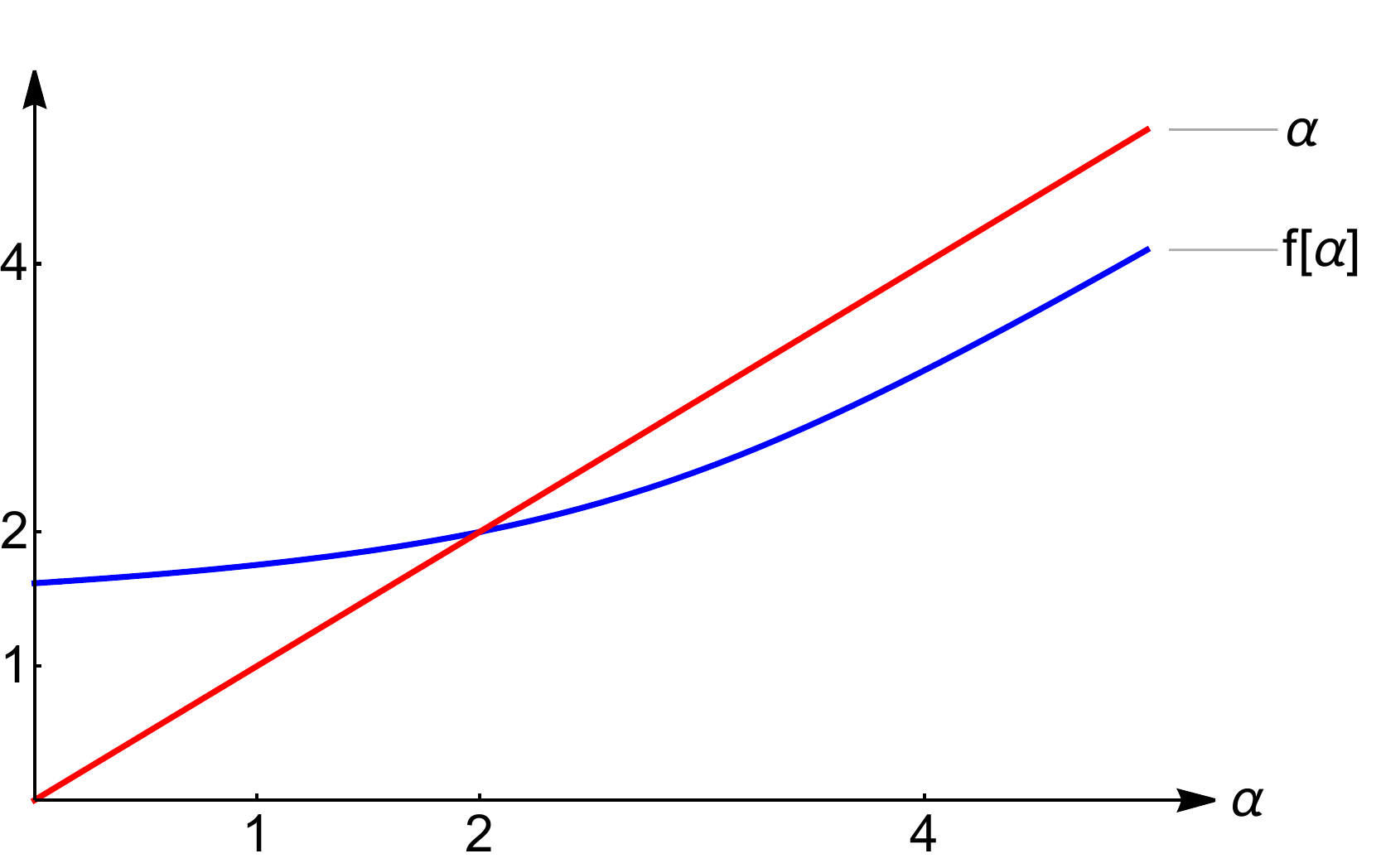}
		\caption{$f(\protect\alpha )$ vs. $\protect\alpha $}
		\label{fig:fava}
	\end{figure}
	From \eqref{eq:falpha} and also from the plot shown in \cref{fig:fava}, we see the following properties%
	\begin{eqnarray}
		f\left( \alpha \right) &=&\alpha \Leftrightarrow \alpha =2;  \label{plott} \\
		0 &\leqslant &\alpha <2:f\left( \alpha \right) >\alpha ; \\
		\alpha &>&2:f\left( \alpha \right) <\alpha ,
	\end{eqnarray}%
	which will be useful later.  $\alpha >1$ (resp.
	$\alpha <1$) corresponds to an over-rotating (resp. under-rotating) BHs,
	whereas $\alpha =0$ corresponds to a non-rotating (i.e., static, RN) BH.

	Moreover, in \cite{Chattopadhyay:2023nfc} it has been further emphasized
	that RFD cannot induce a change in the BH rotation regime; the story goes
	differently for \textit{doubly-extremal} rotating solutions : in \cref{app:anotherrfd} we will show that the (suitably defined) RFD\ \textit{%
		necessarily} maps to over-rotating solutions.
	
	\subsection{RFD on \textit{Doubly-Extremal} BHs}\label{sec:no_or_KNBH}
	
	\label{app:anotherrfd} 
	After the treatment of \cite{Astefanesei:2006dd}, a rotating
	BH with angular momentum $\tilde{J}$ and non-rotating (i.e., static) entropy
	$S_{0}\equiv S_{0}(Q)$, which is \textit{doubly-extremal}, i.e. it is
	coupled to \textit{constant} scalar fields, has a Bekenstein-Hawking entropy%
	\begin{equation}
		S_{BH}^{\text{doubly-extr.}}=\sqrt{S_{0}^{2}\left( Q\right) +4\pi ^{2}\tilde{%
				J}^{2}}.  \label{fthat}
	\end{equation}
	For convenience, we absorb $2\pi$ in the definition of the angular momentum as $2\pi \tilde{J}%
	=:J$, hence the entropy can be written as 
	\begin{equation}
		S_{BH}^{\text{doubly-extr.}}=\sqrt{S_{0}^{2}\left( Q\right) +J^{2}},
		\label{KNenEM}
	\end{equation}%
	
	where $S_{0}\left( Q\right) $ is the (purely $Q$-dependent) non-rotating
	(i.e., static) Bekenstein-Hawking entropy (i.e., the entropy of the RN
	extremal BH obtained by letting $J\rightarrow 0^{(+)}$). The functional form
	of the BH entropy (\ref{KNenEM}), which contains the sum of two squares
	under the square root, has not been treated in the analysis given in \cite%
	{Chattopadhyay:2023nfc}; here, we fill this gap and, along the same lines of
	reasoning of \cite{Chattopadhyay:2023nfc}, we briefly discuss the
	possibility to define a RFD for rotating, doubly-extremal BHs (which we will
	define RFD$^{\text{doubly-extr.}}$).
	
	First of all, we observe that, as for the analysis given in \cite%
	{Chattopadhyay:2023nfc}, a na\"{\i}ve definition of RFD with unvaried $J$ is
	not possible. Indeed, one would obtain the following transformation of the
	e.m. charge symplectic vector :
	\begin{equation}
		\text{RFD}^{\text{doubly-extr.}}:Q\longmapsto \hat{Q}(Q,J):=\Omega \frac{%
			\partial S_{BH}^{\text{doubly-extr.}}}{\partial Q}=\frac{S_{0}\left(
			Q\right) }{\sqrt{S_{0}^{2}\left( Q\right) +J^{2}}}\Omega \frac{\partial
			S_{0}(Q)}{\partial Q}.  \label{canfEM}
	\end{equation}%
	Thus, the condition of invariance of the BH entropy under RFD$^{\text{%
			doubly-extr.}}$ reads%
	\begin{equation}
		S_{BH}^{\text{doubly-extr.}}\left( Q,J\right) =S_{BH}^{\text{doubly-extr.}%
		}\left( \hat{Q}(Q,J),J\right) \Leftrightarrow
		S_{0}^{2}J^{2}(J^{2}+2S_{0}^{2})=0,
	\end{equation}%
	whose only possible solution is $J=0$, thus trivializing the RFD$^{\text{%
			doubly-extr.}}$ to FD itself.
	
	\addtocontents{toc}{\protect\setcounter{tocdepth}{1}}
	
	Therefore, we proceed to consider a transformation of the angular momentum $J
	$ under the RFD, as well. The corresponding transformation rule of $Q$ reads
	\begin{equation}
		\text{RFD}^{\text{doubly-extr.}}:Q\longmapsto \hat{Q}(Q,J+\delta J):=\Omega
		\frac{\partial S_{BH}^{\text{doubly-extr.}}\left( Q,J+\delta J\right) }{%
			\partial Q}=\frac{S_{0}}{\sqrt{S_{0}^{2}+(J+\delta J)^{2}}}\Omega \frac{%
			\partial S_{0}(Q)}{\partial Q}.  \label{RFDEM}
	\end{equation}%
	This implies that the condition of invariance of the BH entropy under RFD$^{%
		\text{doubly-extr.}}$, \textit{i.e.}%
	\begin{equation}
		S_{BH}^{\text{doubly-extr.}}\left( Q,J\right) =S_{BH}^{\text{doubly-extr.}%
		}\left( \hat{Q}(Q,J+\delta J),J+\delta J\right) ,  \label{tthat}
	\end{equation}%
	holds \textit{iff} $\delta J$ satisfies the following inhomogeneous
	algebraic equation of degree six :
	\begin{equation}
		b_{1}(\delta J)^{6}+b_{2}(\delta J)^{5}+b_{3}(\delta J)^{4}+b_{4}(\delta
		J)^{3}+b_{5}(\delta J)^{2}+b_{6}(\delta J)+b_{7}=0, 
		\label{sexticextrkn}
	\end{equation}%
	with
	\begin{eqnarray}
		b_{1} &=&-1,  \notag \\
		b_{2} &=&-6J,  \notag \\
		b_{3} &=&-14J^{2}-S_{0}^{2},  \notag \\
		b_{4} &=&-4J(4J^{2}+S_{0}^{2}),  \notag \\
		b_{5} &=&-9J^{4}-4J^{2}S_{0}^{2}+S_{0}^{4},  \notag \\
		b_{6} &=&-2J(J^{4}+S_{0}^{4}),  \notag \\
		b_{7} &=&J^{2}S_{0}^{2}(J^{2}+2S_{0}^{2}).
	\end{eqnarray}%
	In order to find out two rotating doubly-extremal BHs with different angular
	momenta but the same entropy, having their dyonic charges related by RFD (%
	\ref{RFDEM}), we need to look for a real solution of \eqref{sexticextrkn}%
	, namely for $\delta J=\delta J(J,S_{0})\in \mathbb{R}$ such that the
	transformed angular momentum is (cf. the start of Sec. \ref{sec:cardy})%
	\begin{equation}
		\hat{J}:=J+\delta J\in \mathbb{R}^{+}.  \label{def-J-hat}
	\end{equation}
	
	\subsubsection*{\label{sec:golden}Non-rotating limit : the spurious,
		\textquotedblleft golden\textquotedblright\ branch of doubly-extremal BHs}
	
	Before dealing with a detailed analysis of the roots of the algebraic
	equation \eqref{sexticextrkn}, let us investigate the limit $J\rightarrow
	0^{(+)}$ of the set of solutions to (\ref{sexticextrkn}). In this limit, $%
	\hat{J}\rightarrow \delta J^{(+)}\in \mathbb{R}^{+}$, and Eq. (\ref%
	{sexticextrkn}) reduces to
	\begin{equation}
		(\delta J)^{6}+S_{0}^{2}(\delta J)^{4}-S_{0}^{4}(\delta J)^{2}=0,
		\label{eq:sexticJ0}
	\end{equation}%
	which consistently admits the solution $\delta J=0$. This is no surprise,
	since in the $J\rightarrow 0^{(+)}$ limit, RFD$^{\text{doubly-extr.}}$
	simplifies down to its usual, non-rotating definition (namely, to FD, \cite%
	{Ferrara:2011gv}).
	
	Interestingly, other two real solutions to the cubic algebraic homogeneous
	equation (in $(\delta J)^{2}$) (\ref{eq:sexticJ0}) exist. In fact, Eq. %
	\eqref{eq:sexticJ0} can be solved by two more real roots\footnote{%
		There exist also two purely imaginary roots with $\delta
		J^{2}=-S_{0}^{2}\left( {\frac{1+\sqrt{5}}{2}}\right) $, which we ignore.}

\begin{equation}
	\delta J_{\pm }:=\pm \sqrt{\phi -1}\,S_{0}\left( Q\right) =\pm \frac{1}{%
		\sqrt{\phi }}S_{0}\left( Q\right) ,  \label{golden}
\end{equation}%
with $\phi :={\frac{1+\sqrt{5}}{2}}$ being the so-called \textit{golden ratio%
} (see e.g. \cite{TheFibonacciSequenceandtheGoldenRatio}).
Since $\hat{J}\rightarrow \delta J^{(+)}\in \mathbb{R}^{+}$, only $\delta
J_{+}$ in (\ref{golden}) has a sensible physical meaning. Correspondingly,
within the $J\rightarrow 0^{+}$ (i.e., non-rotating) limit, the unique
non-vanishing, physically sensible solution for the angular momentum
transformation reads
\begin{equation}
	\hat{J}_{\text{golden}}^{\text{doubly-extr.}}:=\delta J_{+}=\sqrt{\phi -1}%
	S_{0}\left( Q\right) =\sqrt{\frac{-1+\sqrt{5}}{2}}S_{0}(Q).
	\label{eq:delJat0}
\end{equation}%
This analysis shows the existence, in the limit $J\rightarrow 0^{+}$, of a
spurious, \textquotedblleft golden\textquotedblright\ branch RFD$%
_{J\rightarrow 0^{+}\text{,~golden}}^{\text{doubly-extr.}}$ of RFD$^{\text{%
		doubly-extr.}}$, which in the non-rotating limit $J\rightarrow 0^{(+)}$ does
\textit{not} reduce to the usual FD, but rather it allows to map a
non-rotating doubly-extremal BH to an under-rotating (stationary)
double-extremal one; this can be depicted as
\begin{equation}
	\text{RFD}_{J\rightarrow 0^{+}\text{,~golden}}^{\text{doubly-extr.}}:%
	\underset{\text{static~doubly-extremal BH}}{\left\{
		\begin{array}{l}
			S=S_{0}\left( Q\right) \\
			J=0%
		\end{array}%
		\right. }~\longrightarrow \underset{\text{under-rotating~doubly-extremal~BH}}%
	{\left\{
		\begin{array}{l}
			S=S_{BH}^{\text{doubly-extr.}}\left( \hat{Q}_{\text{golden}}^{\text{%
					doubly-extr.}},\hat{J}_{\text{golden}}^{\text{doubly-extr.}}\right) \\
			J=\hat{J}_{\text{golden}}^{\text{doubly-extr.}}%
		\end{array}%
		\right. }~,  \label{depict}
\end{equation}%
where $\hat{J}_{\text{golden}}^{\text{doubly-extr.}}$ is defined by (\ref%
{eq:delJat0}), $\hat{Q}_{\text{golden}}^{\text{doubly-extr.}}$ is defined by%
\begin{eqnarray}
	\hat{Q}_{\text{golden}}^{\text{doubly-extr.}} &:&=\Omega \frac{\partial
		S_{BH}^{\text{doubly-extr.}}\left( Q,\hat{J}_{\text{golden}}^{\text{%
				doubly-extr.}}\right) }{\partial Q}=\frac{S_{0}\left( Q\right) }{\sqrt{%
			\left( \hat{J}_{\text{golden}}^{\text{doubly-extr.}}\right)
			^{2}+S_{0}^{2}\left( Q\right) }}\Omega \frac{\partial S_{0}\left( Q\right) }{%
		\partial Q}  \notag \\
	&=&\frac{S_{0}\left( Q\right) }{\sqrt{\left( \phi -1\right)
			S_{0}^{2}(Q)+S_{0}^{2}\left( Q\right) }}\Omega \frac{\partial S_{0}\left(
		Q\right) }{\partial Q}=\frac{1}{\sqrt{\phi }}\Omega \frac{\partial
		S_{0}\left( Q\right) }{\partial Q}  \notag \\
	&=&\sqrt{\phi -1}\Omega \frac{\partial S_{0}\left( Q\right) }{\partial Q},
	\label{Qhat}
\end{eqnarray}%
and RFD$_{J\rightarrow 0^{+}\text{,~golden}}^{\text{doubly-extr.}}$ denotes
such a spurious, \textquotedblleft golden\textquotedblright\ branch of the
non-rotating limit of RFD applied to doubly-extremal BHs, defined by (\ref%
{RFDEM}) :%
\begin{equation}
	\text{RFD}_{J\rightarrow 0^{+}\text{,~golden}}^{\text{doubly-extr.}}:\left\{
	\begin{array}{l}
		Q\rightarrow \hat{Q}_{\text{golden}}^{\text{doubly-extr.}}=\sqrt{\phi -1}%
		\Omega \frac{\partial S_{0}\left( Q\right) }{\partial Q}; \\
		~ \\
		J=0\rightarrow \hat{J}_{\text{golden}}^{\text{doubly-extr.}};%
	\end{array}%
	\right.  \label{RFD-J=0}
\end{equation}%
Note that in the last step of (\ref{Qhat}) we used the crucial property of
the golden ratio $\phi $, namely%
\begin{equation}
	\phi -1=\frac{1}{\phi }.  \label{cruc}
\end{equation}%
Consistently, the total doubly-extremal BH entropy $S_{BH}^{\text{%
		doubly-extr.}}$ is preserved by the map RFD$_{J\rightarrow 0^{+}\text{%
		,~golden}}^{\text{doubly-extr.}}$ (\ref{RFD-J=0}), because%
\begin{eqnarray}
	S_{BH}^{\text{doubly-extr.}}\left( \hat{Q}_{\text{golden}}^{\text{%
			doubly-extr.}},\hat{J}_{\text{golden}}^{\text{doubly-extr.}}\right) &=&\sqrt{%
		S_{0}^{2}\left( \hat{Q}_{\text{golden}}^{\text{doubly-extr.}}\right) +\left(
		\hat{J}_{\text{golden}}^{\text{doubly-extr.}}\right) ^{2}}  \notag \\
	&=&\sqrt{S_{0}^{2}\left( \sqrt{\phi -1}\Omega \frac{\partial S_{0}\left(
			Q\right) }{\partial Q}\right) +\left( \phi -1\right) S_{0}^{2}\left(
		Q\right) }  \notag \\
	&=&\sqrt{\left( \phi -1\right) ^{2}S_{0}^{2}\left( \Omega \frac{\partial
			S_{0}\left( Q\right) }{\partial Q}\right) +\left( \phi -1\right)
		S_{0}^{2}\left( Q\right) }  \notag \\
	&=&\sqrt{\phi ^{2}-\phi }S_{0}\left( Q\right) =S_{0}\left( Q\right) ,
	\label{!}
\end{eqnarray}%
where in the last step the crucial property (\ref{cruc}) has been used
again. Moreover, in achieving (\ref{!}), we have also exploited two crucial
properties of the static (doubly-)extremal BH entropy $S_{0}(Q)$, namely its
homogeneity of degree two in the e.m. charges,%
\begin{equation}
	S_{0}\left( \lambda Q\right) =\lambda ^{2}S_{0}(Q),~\forall \lambda \in
	\mathbb{R},  \label{hom-2-Q}
\end{equation}%
and its invariance under the usual FD \cite{Ferrara:2011gv},%
\begin{equation}
	S_{0}(Q)=S_{0}\left( \Omega \frac{\partial S_{0}\left( Q\right) }{\partial Q}%
	\right) .  \label{FD-inv}
\end{equation}%
Thus, the two doubly-extremal BHs in the l.h.s and r.h.s of (\ref{depict})
have the same Bekenstein-Hawking entropy, preserved by the map RFD$%
_{J\rightarrow 0^{+}\text{,~golden}}^{\text{doubly-extr.}}$ (\ref{RFD-J=0}).

\paragraph*{Remark 1}

It is worth remarking that the stationary doubly-extremal BH in the r.h.s.
of (\ref{depict}), namely the image of a doubly-extremal, static BH under
the map RFD$_{J\rightarrow 0^{+}\text{,~golden}}^{\text{doubly-extr.}}$ (\ref%
{RFD-J=0}), is necessarily \textit{under-rotating}. In fact,%
\begin{equation}
	0<\phi -1<1\Rightarrow \hat{J}_{\text{golden}}^{\text{doubly-extr.}}=\sqrt{%
		\phi -1}S_{0}\left( Q\right) <S_{0}\left( Q\right) \Leftrightarrow \hat{J}_{%
		\text{golden}}^{\text{doubly-extr.}}-S_{0}^{2}\left( Q\right) <0,
\end{equation}%
which pertains to the under-rotating case. This is to be contrasted with
what holds for extremal but not doubly-extremal rotating BHs; cf. remark 1
in Sec. 3.1 of \cite{Chattopadhyay:2023nfc}.

\paragraph*{Remark 2}

The comparison with the spurious, \textquotedblleft
golden\textquotedblright\ branch of the RFD applied to (under- or
over-)rotating, extremal (but not doubly-extremal) BHs, discovered in \cite%
{Chattopadhyay:2023nfc}, is particularly relevant. From Eqs. (\ref%
{eq:delJat0}) and (\ref{Qhat}), the following relations hold:%
\begin{eqnarray}
	\hat{J}_{\text{golden}}^{\text{doubly-extr.}} &=&\sqrt{\frac{\phi -1}{\phi }}%
	\hat{J}_{\text{golden}}=\frac{\hat{J}_{\text{golden}}}{\phi }; \\
	\hat{Q}_{\text{golden}}^{\text{doubly-extr.}} &=&-\sqrt{\frac{\phi -1}{\phi }%
	}\hat{Q}_{\text{golden}}=-\frac{\hat{Q}_{\text{golden}}}{\phi },
\end{eqnarray}%
where $\hat{J}_{\text{golden}}$ and $\hat{Q}_{\text{golden}}$ are
respectively given by Eqs. (3.8) and (3.10) of \cite{Chattopadhyay:2023nfc}.

\paragraph*{Remark 3}

In a sense, the use of the spurious, \textquotedblleft
golden\textquotedblright\ branch RFD$_{J\rightarrow 0^{+}\text{,~golden}}^{%
	\text{doubly-extr.}}$ (\ref{RFD-J=0}) of the map RFD (\ref{RFDEM}) can be
regarded as a kind of solution-generating technique, which generates an
under-rotating, stationary doubly-extremal BH from a non-rotating, static
BH, while keeping the Bekenstein-Hawking entropy fixed. Indeed, while both
such BHs are asymptotically flat, their near-horizon geometry changes under
RFD$_{J\rightarrow 0^{+}\text{,~golden}}^{\text{doubly-extr.}}$ :%
\begin{equation}
	\underset{\text{static~doubly-extremal BH~\cite{Kunduri:2007vf}}}{%
		AdS_{2}\otimes S^{2}}~\overset{\text{RFD}_{J\rightarrow 0^{+}\text{,~golden}%
		}^{\text{doubly-extr.}}}{\longrightarrow }\underset{\text{%
			under-rotating~doubly-extremal~BH \cite{Bardeen:1999px}}}{AdS_{2}\otimes
		S^{1}}~.  \label{depict-nhg}
\end{equation}%
This is totally analogous to what holds for extremal (but not
doubly-extremal) rotating BHs; see remark 2 in Sec. 3.1 of \cite%
{Chattopadhyay:2023nfc}.

\paragraph*{Remark 4}

The expression of the Bekenstein-Hawking entropy $S_{BH}^{\text{doubly-extr.}%
}$ of doubly-extremal (stationary) rotating BHs, given by (\ref{that}) or
equivalently by (\ref{KNenEM}), is suggestive of a representation of the two
fundamental quantities $S_{0}$ and $J$ characterizing such BHs in terms of
elements of the \textit{complex} numbers $\mathbb{C}$, i.e. respectively as%
\begin{equation}
	Z:=S_{0}+\mathbf{i}J\in \mathbb{C}~\Rightarrow S_{BH}^{\text{doubly-extr.}%
	}=\left\vert Z\right\vert :=\sqrt{Z\bar{Z}}=\sqrt{S_{0}^{2}+J^{2}},
\end{equation}%
where the bar stands for the complex conjugation.

Thus, any map acting on $S_{0}$ and $J$ of a doubly-extremal rotating
(stationary) BH, as the RFD map (\ref{RFDEM}) itself, can be represented as
acting on the Argand-Gauss plane $\mathbb{C}\left[ S_{0},J\right] $ (or,
equivalently, $\mathbb{C}\left[ J,S_{0}\right] $) itself. Since complex
numbers are a field (because they do not contain divisors of $0$), there may
exist \textquotedblleft wild\textquotedblright\ automorphisms \cite%
{Kestelman} of $\mathbb{C}$, depending on the so-called `axiom of choice'.
The four discrete fundamental automorphisms are $\mathbb{I}$, $-\mathbb{I}$,
$\mathbf{C}$ and $-\mathbf{C}$, where $\mathbb{I}$ and $\mathbf{C}$
respectively denote the identity map and the aforementioned conjugation map;
interestingly, since $S_{0}\in \mathbb{R}_{0}^{+}$ and $J\in \mathbb{R}^{+}$%
, none (but, trivially, $\mathbb{I}$) of such discrete automorphisms are
physically allowed, and the physically sensible quadrant of $\mathbb{C}\left[
S_{0},J\right] $ is only the first one.

In particular, the RFD non-linear map (\ref{RFDEM}) map acts on any $Z\in
\mathbb{C}\left[ S_{0},J\right] $ as a norm-preserving transformation,
because, by definition, it preserves the Bekenstein-Hawking entropy $S_{BH}^{%
	\text{doubly-extr.}}$, as given by the defining condition (\ref{KNenEM}).
Thus, considering a (stationary, asymptotically flat) doubly-extremal
rotating BH with Bekenstein-Hawking entropy $S_{BH}^{\text{doubly-extr.}}=%
\mathcal{S}\in \mathbb{R}^{+}$, its RFD-dual extremal BH will belong to the
arc of the circle of squared radius $\left\vert Z\right\vert ^{2}=\mathcal{S}%
^{2}$ within the first quadrant of $\mathbb{C}\left[ S_{0},J\right] $.

In the non-rotating limit, in light of the treatment above, the action of
the RFD map (\ref{RFDEM}) on a non-rotating doubly-extremal BH (represented
as a point of the $S_{0}$ axis - with its origin excluded - in $\mathbb{C}%
\left[ S_{0},J\right] $, thus with coordinates $\left( S_{0},0\right) $) may
be nothing but the identity $\mathbb{I}$ (in the usual branch of RFD$%
_{J\rightarrow 0^{+}}^{\text{doubly-extr.}}$) or a point with coordinates $%
\left( \left( \phi -1\right) S_{0},\sqrt{\phi -1}S_{0}\right) =\left( \frac{%
	S_{0}}{\phi },\frac{S_{0}}{\sqrt{\phi }}\right) $ along the arc of the
circle defined as $\mathcal{C}:=\left\{ Z\in \mathbb{C}\left[ S_{0},J\right]
:\left\vert Z\right\vert ^{2}=S_{0}^{2}\right\} $ (in the spurious,
\textquotedblleft golden\textquotedblright\ branch RFD$_{J\rightarrow 0^{+}%
	\text{,~golden}}^{\text{doubly-extr.}}$ discussed above). Again, this is to
be contrasted with remark 3 in Sec. 3.1 of \cite{Chattopadhyay:2023nfc}.


\subsubsection*{Analytical solution}

\begin{figure}[h]
	\centering
	\includegraphics[width=0.75\textwidth]{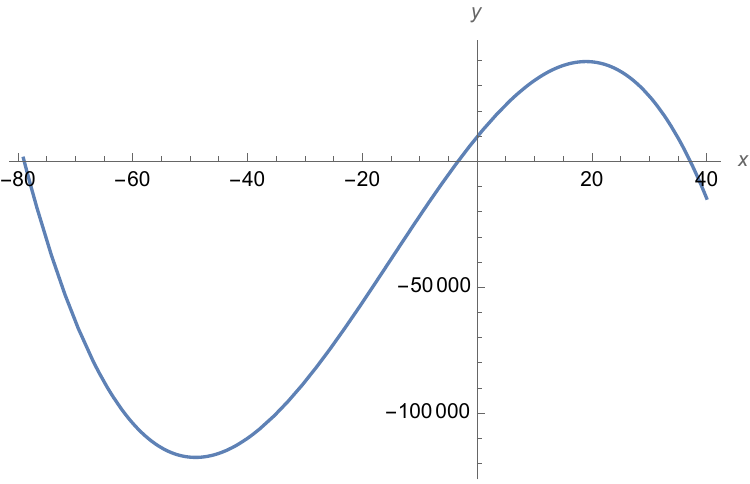}
	\caption{Typical plot of $g(x)$ for $J\neq 0$ .}
	\label{fig:extrkn}
\end{figure}

By recalling the definition (\ref{def-J-hat}), the sextic equation (\ref{sexticextrkn}) can be rewritten as
\begin{equation}
	-\hat{J}^{6}+\hat{J}^{4}(J^{2}-S_{0}^{2})+\hat{J}%
	^{2}S_{0}^{2}(S_{0}^{2}+2J^{2})+J^{2}S_{0}^{4}=0,
\end{equation}%
and by setting $x:=\hat{J}^{2}$, one obtains the following cubic
inhomogeneous equation in $x$ :
\begin{equation}
	-x^{3}+x^{2}(J^{2}-S_{0}^{2})+xS_{0}^{2}(S_{0}^{2}+2J^{2})+J^{2}S_{0}^{4}=0.
	\label{cubicEM}
\end{equation}%
As discussed \textit{e.g. }in \cite{Chattopadhyay:2023nfc} (see also Refs.
therein), the turning points of the parametric curve
\begin{equation}
	g(x):=-x^{3}+x^{2}(J^{2}-S_{0}^{2})+xS_{0}^{2}(S_{0}^{2}+2J^{2})+J^{2}S_{0}^{4}
\end{equation}%
tell us about the root of the equation \eqref{cubicEM}. By solving $\frac{%
	dg(x)}{dx}=0$, we find the two turning points $T_{i}=(x_{i},g(x_{i})),i=1,2$
with coordinates%
\begin{eqnarray}
	T_{1} &=&(-S_{0}^{2},-S_{0}^{6}), \\
	T_{2} &=&\left( \frac{1}{3}(2J^{2}+S_{0}^{2}),\frac{\mathcal{B}}{27}\right) ,
	\\
	\mathcal{B} &:&=\left( 4(J^{2}+2S_{0}^{2})^{3}-27S_{0}^{6}\right)
	=(4J^{6}+24J^{4}S_{0}^{2}+48J^{2}S_{0}^{4}+5S_{0}^{6})>0.
\end{eqnarray}%
Hence, the discriminant of $g(x)$ is $\Delta =S_{0}^{6}\mathcal{B}>0$. Thus,
the equation $g(x)=0$ always has three real roots. Note that $%
g(0)=J^{2}S^{0} $ is always on the positive $y\equiv g(x)$ axis. The two
turning points $T_{1}$ and $T_{2}$ lie in the third quadrant and first
quadrant respectively in the $x$ \textit{vs.} $y=g(x)$ plot. This implies
that $g(x)$ has only one positive real root which is physically sensible,
namely $x=\hat{J}^{2}$, always a positive and real quantity. Thus, acting
with RFD$_{J\rightarrow 0^{+}}^{\text{doubly-extr.}}$ on a doubly-extremal
KN BH with charge $Q$ and angular momentum $J$, one obtains another
doubly-extremal rotating unique BH with charge $\hat{Q}(Q,\hat{J})$ and
angular momentum $\hat{J}$. A typical plot of $g(x)$ \textit{vs.} $x$ (for $%
J=2,S_{0}=7$) is shown in figure \ref{fig:extrkn}.

\subsubsection*{Numerical solution}

We can also solve the equation \eqref{sexticextrkn} numerically, and see that there exist only two real solutions
for $\delta J$, namely $\delta J_{1}$ and $\delta J_{2}$. Then, these two
real solutions can be numerically analyzed by choosing some positive and
real values for $J$ and $S_{0}$ : it is easy to realize that there is only
one solution for $\delta J$ that makes $\hat{J}:=J+\delta J$ always
positive. For example, with $S_{0}=2$, we plot the $J+\delta J_{i}$ \textit{%
	vs.} $J_{i}$ ($i=1,2$) curves as follows
\begin{figure}[tbph]
	\begin{subfigure}{0.48\textwidth}
		\centering
		\includegraphics[width=\textwidth]{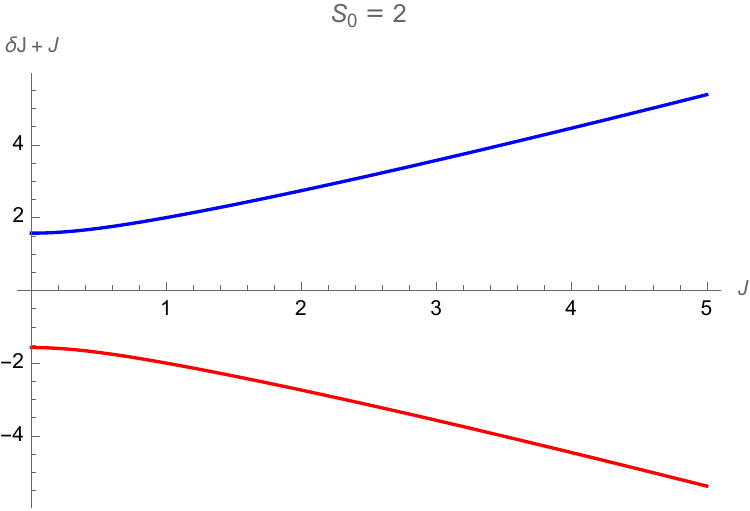}
		\caption{Fixed Entropy}
	\end{subfigure}\hfill
	\begin{subfigure}{0.48\textwidth}
		\centering
		\includegraphics[width=\textwidth]{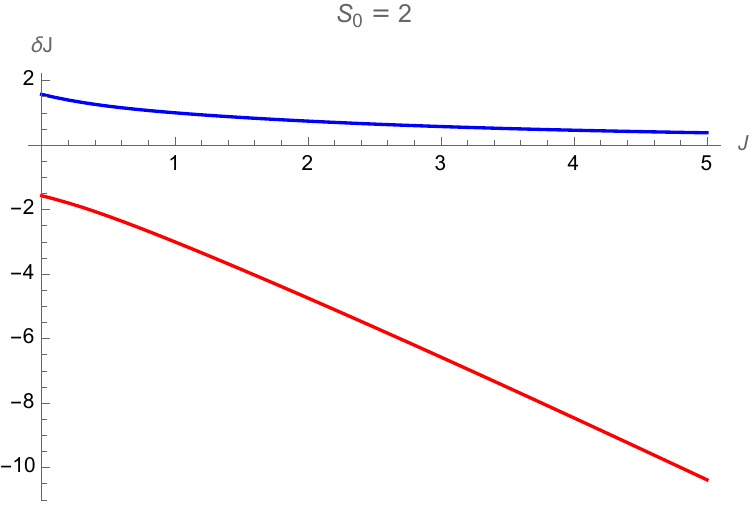}
		\caption{Fixed entropy}
	\end{subfigure}
	\caption{Two real roots of (\protect\ref{sexticextrkn})}
	\label{fig:numerics}
\end{figure}

\subsubsection*{Exact solution}

A convenient way of solving \eqref{cubicEM} is to write it in the \textit{%
	depressed form}
\begin{equation}
	\mathbf{t}^{3}+\mathbf{p}\mathbf{t}+\mathbf{q}=0,  \label{depressed}
\end{equation}%
with $\mathbf{t}:=x-\frac{1}{3}(J^{2}-S_{0}^{2}),\quad \mathbf{p}:=-\frac{1}{%
	3}(J^{2}+2S_{0}^{2})^{2},\quad \mathbf{q}:=(\frac{S_{0}^{6}}{2}-\frac{%
	\mathcal{B}}{54})$. Therefore, the solution to \eqref{depressed} is given by
\begin{equation}\label{eq:cubicrt}
	\mathbf{t}=\sqrt[3]{ -\frac{\mathbf{q}}{2}+\sqrt{\frac{\mathbf{q}^{2}}{4}+\frac{%
				\mathbf{p}^{3}}{27}}}+\sqrt[3]{ -\frac{\mathbf{q}}{2}-%
		\sqrt{\frac{\mathbf{q}^{2}}{4}+\frac{\mathbf{p}^{3}}{27}}}.
\end{equation}%
One should note that in the above expression the operations "$\sqrt{}$" and "$\sqrt[3]{}$" represents the principle value of the expression making \eqref{eq:cubicrt} always real. Plugging back the definitions of $\mathbf{t},\mathbf{p}$ and $\mathbf{q}$, one finds that the positive real root can be written as
\begin{equation}
	\hat{J}^{2}=x={1\over 3}(J^2-S_0^2)-\omega\left({S_0^3\over 2}+\sqrt{\mathcal{-B}\over 108}\right)^{2\over 3}-\omega^2\left( {S_0^3\over 2}-\sqrt{\mathcal{-B}\over 108}\right)^{2\over 3}
\end{equation}
where $\omega(=-{1\over 2}-i{\sqrt{3}\over 2})$ and $\omega^2$ are the two non trivial cubic roots of unity with the relation that $\omega^2=1/\omega$.
Similar to the case of extremal black hole, defining $%
\alpha :=J^{2}/S_{0}^{2}$, this formula can be recast as follows :
\begin{eqnarray}\label{hja}
	\hat{J}^{2}&=&\frac{S_{0}^{2}}{3}(\alpha -1)-\omega{S_0^2\over 2^{2/3}}\left(1+\sqrt{1-{4\over 27}(\alpha+2)^3}\right)^{2\over 3}-\omega^2{S_0^2\over 2^{2/3}}\left(1-\sqrt{1-{4\over 27}(\alpha+2)^3}\right)^{2\over 3}\notag\\&=&:S_{0}^{2}h(\alpha )
\end{eqnarray}
where
\begin{equation}\label{eq:ha}
	h(\alpha)={1\over 3}(\alpha-1)-{\omega\over 2^{2/3}}\left(1+\sqrt{1-{4\over 27}(\alpha+2)^3}\right)^{2\over 3}-{\omega^2\over 2^{2/3}}\left(1-\sqrt{1-{4\over 27}(\alpha+2)^3}\right)^{2\over 3}
\end{equation}
to be compared with $f\left( \alpha \right) $, defined in 
(%
\ref{eq:falpha}).

\begin{figure}[t]
	\centering
	\includegraphics[width=0.75\linewidth]{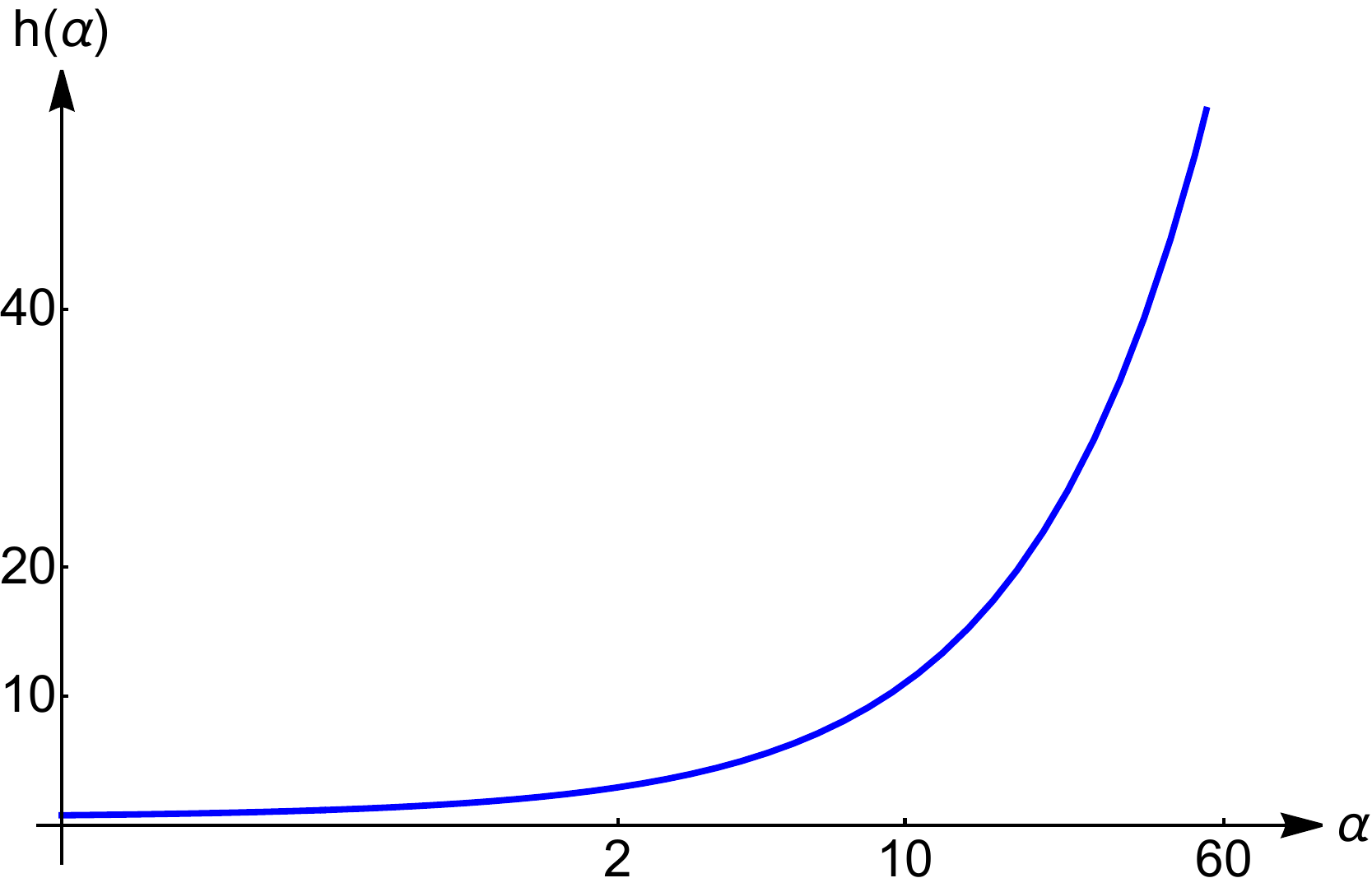}
	\caption{ Semi-log plot of $h(\protect\alpha )$ \textit{vs.} $\protect\alpha $}
	\label{fig:hbeta}
\end{figure}

We can express the RFD$^{\text{doubly-extr.}}$-transformed entropy as
\begin{eqnarray}
	S_{BH}^{\text{doubly-extr.}}(\hat{Q}(Q,\hat{J}),\hat{J}) &:&=\sqrt{%
		S_{0}^{2}\left( \hat{Q}(Q,\hat{J})\right) +\hat{J}^{2}}=\sqrt{%
		S_{0}^{2}\left( \frac{S_{0}}{\sqrt{S_{0}^{2}+\hat{J}^{2}}}\Omega \frac{%
			\partial S_{0}(Q)}{\partial Q}\right) +\hat{J}^{2}}  \notag \\
	&=&\sqrt{\frac{S_{0}^{4}}{\left( S_{0}^{2}+\hat{J}^{2}\right) ^{2}}%
		S_{0}^{2}\left( \Omega \frac{\partial S_{0}(Q)}{\partial Q}\right) +\hat{J}%
		^{2}}=\sqrt{\frac{S_{0}^{6}}{\left( S_{0}^{2}+\hat{J}^{2}\right) ^{2}}+\hat{J%
		}^{2}}  \notag \\
	&=&\sqrt{\Tilde{S_{0}}^{2}(Q,\hat{J})+\hat{J}^{2}},
\end{eqnarray}%
where%
\begin{equation}
	\Tilde{S_{0}}(Q,\hat{J}):=\frac{S_{0}^{3}(Q)}{S_{0}^{2}(Q)+\hat{J}^{2}}%
	\overset{\text{(\ref{eq:ha})}}{=}\frac{S_{0}(Q)}{1+h(\alpha )}.  \label{x}
\end{equation}%
Note that,consistently, $\Tilde{S_{0}}(Q,\hat{J})$ is nothing but the RFD$^{%
	\text{doubly-extr.}}$-transformed non-rotating BH entropy%
\begin{eqnarray}
	RFD^{\text{doubly-extr.}} &:&S_{0}(Q)\mapsto S_{0}\left( \hat{Q}\right)
	=S_{0}\left( \frac{S_{0}}{\sqrt{S_{0}^{2}+\hat{J}^{2}}}\Omega \frac{\partial
		S_{0}(Q)}{\partial Q}\right) \\
	&=&\frac{S_{0}^{2}}{S_{0}^{2}\left( Q\right) +\hat{J}^{2}}S_{0}\left( \Omega
	\frac{\partial S_{0}(Q)}{\partial Q}\right) =\frac{S_{0}^{3}}{%
		S_{0}^{2}\left( Q\right) +\hat{J}^{2}}=\Tilde{S_{0}}(Q,\hat{J}),  \notag
\end{eqnarray}%
and thus the real positive parameter%
\begin{equation}
	\hat{\alpha}:=\frac{\hat{J}^{2}}{\Tilde{S}_{0}^{2}}  \label{def-alpha-hat}
\end{equation}%
has the physically sensible interpretation of RFD$^{\text{doubly-extr.}}$%
-transformed $\alpha $. Of course, the RFD$^{\text{doubly-extr.}}$%
-invariance of the doubly-extremal KN BH entropy (\ref{tthat}) implies that%
\begin{equation}
	S_{BH}^{\text{doubly-extr.}}(\hat{Q}(Q,\hat{J}),\hat{J})=S_{BH}^{\text{%
			doubly-extr.}}(Q,J)=\sqrt{S_{0}^{2}+J^{2}}=S_{0}\sqrt{1+\alpha }.
\end{equation}%
Finally, Eqs. (\ref{x}) and (\ref{x}), together with definition (\ref%
{def-alpha-hat}), yield that%
\begin{equation}
	\hat{\alpha}=h(\alpha )(1+h(\alpha ))^{2},  \label{ttthat}
\end{equation}%
whereas (\ref{def-alpha-hat}) and (\ref{ttthat}) imply%
\begin{equation}
	\Tilde{S_{0}}\sqrt{1+\hat{\alpha}}=S_{0}\sqrt{1+\alpha }\overset{\text{(\ref%
			{eq:ha})}}{\Leftrightarrow }\frac{\sqrt{1+\hat{\alpha}}}{1+h(\alpha )}=\sqrt{%
		1+\alpha },
\end{equation}%
which, by virtue of (\ref{ttthat}), allows one to determine an equation
satisfies by $h(\alpha )$ itself\footnote{
	An analogous equation for $f(\alpha )$ defined in (\ref{eq:falpha}) can be
	derived from the treatment given in Sec. 7 of \cite{Chattopadhyay:2023nfc}:
	\begin{equation}\label{eq:frel}
		(1-f(\alpha))^2(1-\alpha+f(\alpha))=1.
	\end{equation}
	Actually, $f(\alpha)$ resp. $h(\alpha)$ given by equations resp. (\ref{eq:falpha}) \& (\ref{eq:ha}), are the unique physically sound solution of the functional equation resp. (\ref{eq:hrel}) \& (\ref{eq:frel}). (ie, they are the only solutions implying $\hat{J}$ real and positive.)}:%

\begin{eqnarray}\label{eq:hrel}
	1+h\left( \alpha \right) &=\sqrt{\frac{1+\hat{\alpha}}{1+\alpha }}\overset{%
		\text{(\ref{ttthat})}}{=}&\sqrt{\frac{1+h(\alpha )(1+h(\alpha ))^{2}}{%
			1+\alpha }};  \notag \\
	&\Updownarrow&  \notag \\
	\left( 1+\alpha -h\left( \alpha \right) \right) &\left( 1+h\left( \alpha
	\right) \right) ^{2}=1,&
\end{eqnarray}%
where the \textit{iff} (\textit{i.e.}, \textquotedblleft $\Updownarrow $%
\textquotedblright ) implication holds because $h\left( \alpha \right)
>0~\forall \alpha \geqslant 0$, as evident from the corresponding plot shown in \cref{fig:hbeta}.

\subsubsection*{$\hat{\protect\alpha}>1$ always}

\begin{figure}[h]
	\centering
	\includegraphics[width=0.7\linewidth]{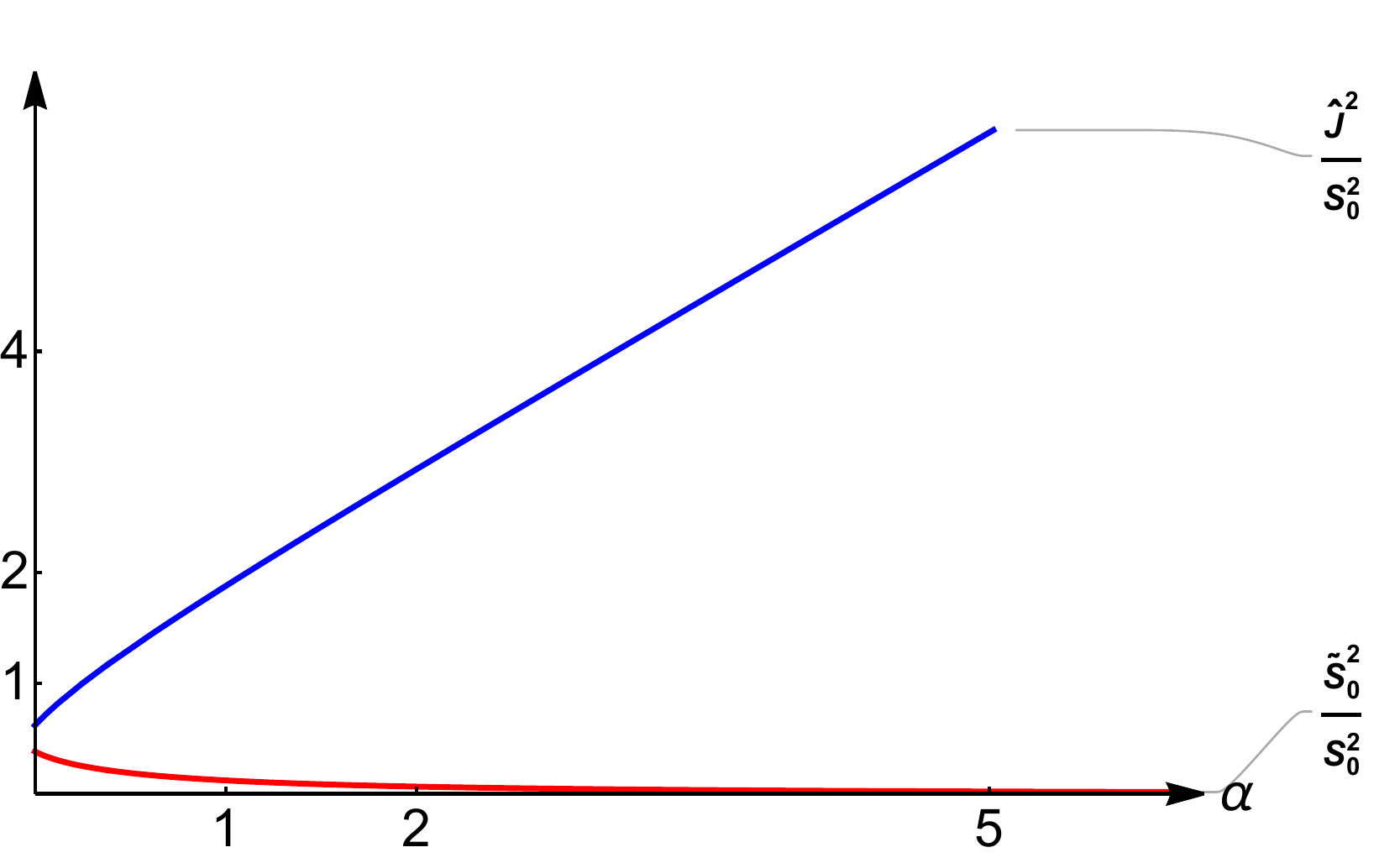}
	\caption{Any RFD-tranformed doubly-extremal KN BH is always over-rotating}
	\label{fig:hbeta1}
\end{figure}

At $\alpha =1\Leftrightarrow J^{2}=S_{0}^{2}$, it holds that $\hat{\alpha}%
\approx 15.58\Rightarrow \hat{J}^{2}>\Tilde{S}_{0}^{2}$. On the other hand, $%
\hat{\alpha}=1\Leftrightarrow \hat{J}^{2}=\Tilde{S}_{0}^{2}\Leftrightarrow
h(\alpha )(1+h(\alpha ))^{2}=1$, implying that $h(\alpha )\approx 0.4656$.
Therefore, if $h(\alpha )<0.4656$, then $\hat{\alpha}<1$, and if $h(\alpha
)>0.4656$, then $\hat{\alpha}>1$. One can check that $h(0)\approx 0.62$, and
also that $h(\alpha )$ is a monotonically increasing function (from graph %
\ref{fig:hbeta}). Hence, for any value of $\alpha \geqslant 0$ it holds that
$\hat{\alpha}>1$. This means that, regardless the under-rotating ($%
0\leqslant \alpha <1$) or over-rotating ($\alpha >1$) regime of (stationary)
rotation of the starting doubly-extremal KN BH, the RFD-transformed
doubly-extremal KN BH is always over-rotating : $\hat{\alpha}>1$. Again,
this is clarified in \cref{fig:hbeta1}.

\addtocontents{toc}{\protect\setcounter{tocdepth}{2}}

\section{Kerr/CFT correspondence and its extensions}

\label{sec:KNBH} The Kerr/CFT correspondence \cite{Guica:2008mu}
establishes a duality between a four-dimensional, asymptotically flat,
extremal Kerr BH and a two-dimensional (thermal) CFT. This duality arises
from the realization that the near-horizon extremal Kerr (NHEK) metric is
essentially equivalent to a warped $AdS_{3}$ space. In contrast to the $SL(2,%
\mathbb{R})_{R}\times SL(2,\mathbb{R})_{L}$ isometry group of $AdS_{3}$, the
warped $AdS_{3}$ has an isometry group of $SL(2,\mathbb{R})_{R}\times U(1)$.
Following a similar analysis by Brown and Henneaux \cite{Brown:1986nw}, the
asymptotic symmetry group of the extremal Kerr BH has been studied in \cite%
{Guica:2008mu}. With a suitable choice of fall off of the metric at the
boundary, the $U(1)$ symmetry of $SL(2,\mathbb{R})_{R}\times U(1)$ gets
enhanced to the infinite-dimensional Virasoro algebra with a central charge $%
c_{L}=12\tilde{J}$, where $\tilde{J}$ is the angular momentum of the dual
BH. To compute the statistical entropy of this CFT, we need the temperature $%
T_{L}$ which cannot be proportional to the Hawking temperature of the dual
extremal Kerr BH, as that is vanishing. In order to find out $T_{L}$, it
becomes necessary to define a proper quantum vacuum. As the Kerr geometry
lacks a global time-like Killing vector, there is no Hartle-Hawking vacuum.
We need to consider Frolov and Thorne vacuum \cite%
{FT1,Ottewill1,Duffy:2005mz} which is analogous to the Hartle-Hawking vacuum
in the near-horizon region. This leads to the finding the temperature $T_{L}=%
\frac{1}{2\pi }$, while $T_{R}=0$. Hence following the Cardy formula, we
find the entropy of the extremal Kerr BH as
\begin{equation}
	S_{\text{Cardy}}^{\text{Kerr}}={\pi^2\over 3}c_L\,T_L=2\pi \tilde{J}=S_{BH}^{\text{Kerr}%
	},  \label{this2}
\end{equation}%
which thus agrees with the Bekenstein-Hawking entropy for the same.

\subsection{Maxwell fields}

The Kerr/CFT correspondence can be extended to the so-called Kerr-Newman/CFT
correspondence, to include Maxwell fields (and thus, extremal KN BH
solutions), generally also in presence of scalar fields \cite{Haco:2019ggi,
	Chow:2008dp, Compere:2009dp} (see also{\ \cite{Compere:2012jk} for a
	comprehensive review, and the treatment below). An extremal KN BH} is
characterized by its charge $q$ and the angular momentum $\tilde{J}$. The
near-horizon isometry is given by $SL(2,\mathbb{R})\times U(1)$.
The asymptotic symmetry group for the near-horizon of the extremal KN BH
contains two sets of Virasoro algebras with central charges $c_{L}=c_{R}=12%
\tilde{J}$. Introducing the Frolov-Thorne vacuum as in the case of Kerr BH,
we find the temperatures of the two CFTs as $T_{L}=\frac{r_{+}}{2\pi a}-%
\frac{q^{2}}{4\pi Ma}$ and $T_{R}=0$, where $M$ and $r_{+}$ are the mass and
the outer event horizon of the BH with $a=\tilde{J}/M$. Hence, the Cardy
entropy is
\begin{eqnarray}
	S_{\text{Cardy}}^{\text{KN}} &=&{\frac{\pi ^{2}}{3}}(c_{L}T_{L}+c_{R}T_{R})
	\label{eq:cardentropy} \\
	&=&{\frac{\pi ^{2}}{3}}c_{L}T_{L}  \label{eq:cardentropy2} \\
	&=&\pi (2Mr_{+}-q^{2})=S_{BH}^{\text{KN}},  \label{this}
\end{eqnarray}%
with $S_{BH}^{\text{KN}}$ being the Bekenstein-Hawking entropy of the
extremal KN BH. To summarize, after the treatment given in \cite%
{Astefanesei:2006dd} and \cite{Compere:2009dp,Compere:2012jk}, the CFT dual
to an extremal KN BH with angular momentum $\Tilde{J}$ and (electric) charge
$Q$ is endowed with the following central charges and temperatures :
\begin{eqnarray}
	c &=&12\Tilde{J};  \label{tl1} \\
	T_{L} &=&\frac{\sqrt{q^{4}+4\Tilde{J}^{2}}}{4\pi \Tilde{J}}; \\
	T_{R} &=&0.
\end{eqnarray}%
Hence the Bekenstein-Hawking entropy $S_{BH}^{\text{KN}}$ and the Cardy
entropy $S_{\text{Cardy}}^{\text{KN}}\equiv S_{CFT}$ of the dual CFT are
\begin{equation}
	S_{BH}^{\text{KN}}=S_{CFT}=\pi \sqrt{q^{4}+4\Tilde{J}^{2}}.
	\label{pre-this3}
\end{equation}%
As to be expected, the $q\rightarrow 0$ and $\Tilde{J}\rightarrow 0^{(+)}$
limits of this formula respectively yield the extremal Kerr BH entropy (\ref%
{this2}) and the extremal (static) RN entropy $\pi q^{2}$.

\subsection{Scalar fields}

\label{kerrcftother} It has been conjectured and proved that the central
charge of a CFT dual to an extremal, rotating (Kerr or KN) BH which solves
Einstein equations of a theory containing scalars and other fields depends
only on the gravitational field \cite{Compere:2009dp,Compere:2012jk}.
Nevertheless, the Bekenstein-Hawking entropy is still the same as the Cardy
entropy of the dual CFT.

Here, we discuss briefly this further generalization of the Kerr/CFT
correspondence. The gravitational action of our concerned theory is of the
form \cite{Astefanesei:2006dd,Compere:2009dp,Compere:2012jk}
\begin{eqnarray}
	\mathcal{S} &=&\frac{1}{16\pi G_{N}}\int d^{4}x\sqrt{-g}\Big(R-\frac{1}{2}%
	f_{AB}(\Phi )\partial _{\mu }\Phi ^{A}\partial ^{\mu }\Phi ^{B}  \notag \\
	&&-V(\Phi )-k_{IJ}(\Phi )F_{\mu \nu }^{I}F^{J\mu \nu }+h_{IJ}(\Phi )\epsilon
	^{\mu \nu \delta \tau }F_{\mu \nu }^{I}F_{\delta \tau }^{J}\Big).
\end{eqnarray}%
Other than gravity, the theory involves scalars $\Phi $ and gauge fields $%
A_{\mu }^{I}$ with field strength $F_{\mu \nu }^{I}$. Here $\mu ,\nu ,$ run
over the space-time indices, $A,B$ denotes the scalar indices and $I,J$
denotes the gauge field indices. Since we are going to consider
asymptotically flat extremal BH solutions, in the rest of the treatment we
will assume $V=0$, so we may be considering (the purely bosonic sector of)
ungauged supergravity, or gauged supergravity with flat Abelian gaugings%
\footnote{%
	This choice is immaterial, since the solutions are exactly the same, as far
	as the supersymmetry (BPS-) preserving properties, and in general fermions,
	are not concerned \cite{Hristov:2012nu}.}, as well. We assume that $%
f_{AB}(\Phi )$ and $k_{IJ}(\Phi )$ are positive definite and maximal rank,
of course. The near-horizon region of any stationary, rotating extremal BH
of this theory has $SL(2,\mathbb{R})\times U(1)$ isometry. Hence, the
near-horizon geometry takes the form \cite%
{Astefanesei:2006dd,Compere:2009dp,Compere:2012jk}
\begin{eqnarray}
	ds^{2} &=&\Gamma (\theta )\big[-r^{2}dt^{2}+\frac{dr^{2}}{r^{2}}+\beta
	(\theta )^{2}d\theta ^{2}+\gamma (\theta )^{2}\left( d\phi +\rho rdt\right)
	^{2}\big]  \notag \\
	&=&\Omega (\theta )^{2}e^{2\Psi (\theta )}\big[-r^{2}dt^{2}+\frac{dr^{2}}{%
		r^{2}}+\beta (\theta )^{2}d\theta ^{2}  \notag \\
	&&+\Omega (\theta )^{-2}e^{-4\Psi (\theta )}\left( d\phi +\rho rdt\right)
	^{2}\big]; \\
	\Phi ^{A} &=&\Phi ^{A}(\theta ),\quad A^{I}=f^{I}(\theta )\left( d\phi +\rho
	rdt\right) +\frac{e^{I}}{\rho }d\phi .
\end{eqnarray}%
The BH entropy is then given by
\begin{equation}
	S_{BH}=\frac{1}{4G_{N}\hbar }\int d\theta d\phi \sqrt{g_{\theta \theta
		}g_{\phi \phi }}=\frac{\pi }{2G_{N}\hbar }\int d\theta \Omega (\theta )\beta
	(\theta ),  \label{bkstein1}
\end{equation}%
with $\Gamma (\theta )>0,\gamma (\theta )\geq 0$. $\Phi ^{A}(\theta
),f^{I}(\theta ),k$ and $e_{I}$ are real parameters who get fixed by solving
the equations of motion.

On the other hand, the Cardy entropy of the dual $2D$ CFT is
\begin{equation}
	S_{CFT}=\frac{\pi ^{2}}{3}cT_{L},
\end{equation}%
with \cite{Compere:2009dp,Compere:2012jk}
\begin{eqnarray}
	&&c=\frac{3\rho }{G_{N}\hbar }\int d\theta \sqrt{g_{\theta \theta }g_{\phi
			\phi }};\label{centralcharge} \\
	&&T_{L}=\frac{1}{2\pi \rho }.\label{lefttemp}
\end{eqnarray}%
Hence, the Cardy entropy is
\begin{equation}
	S_{CFT}=\frac{\pi }{2G_{N}\hbar }\int d\theta \sqrt{g_{\theta \theta
		}g_{\phi \phi }},
\end{equation}%
and it matches with the Bekenstein-Hawking entropy (\ref{bkstein1}).

\subsubsection{Frozen scalars}

\label{carsec2}
As discussed in \ref{app:anotherrfd}, the Bekenstein-Hawking
entropy of rotating, doubly-extremal BHs with fixed scalars at the asymptote is as follows, 

\begin{equation}
	S_{BH}^{\text{doubly-extr.}}=\sqrt{S_{0}^{2}\left( Q\right) +4\pi ^{2}\tilde{%
			J}^{2}}.  \label{that}
\end{equation}%
For doubly-extremal solutions, this expression generalizes the purely
electric, KN formula (\ref{pre-this3}), because $\pi q^{2}$ gets generalized
to a generic, dyonic expression $S_{0}\left( Q\right) $ of the extremal
(static) RN entropy. By the aforementioned KN/CFT correspondence, the dual
CFT is equipped with
\begin{eqnarray}
	c &=&12\tilde{J} \label{fcl}; \\
	T_{L} &=&\frac{\sqrt{S_{0}^{2}\left( Q\right) +4\pi ^{2}\tilde{J}^{2}}}{4\pi
		^{2}\tilde{J}}; \\
	T_{R} &=&0,
\end{eqnarray}%
thus yielding%
\begin{equation}
	S_{CFT}=\sqrt{\left( S_{0}\right) ^{2}+4\pi ^{2}\tilde{J}^{2}}=S_{BH}^{\text{%
			doubly-extr.}}.
\end{equation}%
Completing the analysis done in \cite{Chattopadhyay:2023nfc}, in \cref{app:anotherrfd} we have proved that there exists a unique, Freudenthal dual
BH for doubly-extremal rotating BHs, and we have also found a `spurious' (but
`golden'!) branch of solutions in the non-rotating limit; moreover, we have
also pointed out that the RFD always boosts a rotating
doubly-extremal BH (with any ratio of $S_{0}\left( Q\right) $ and $2\pi
\tilde{J}$) {to an over-rotating regime. }

\subsubsection{Running scalars}

After the treatment of \cite{Astefanesei:2006dd}, for rotating BHs which are
\textit{extremal but not doubly-extremal} (i.e., which are coupled to
running scalar fields), the Bekenstein-Hawking entropy is given by (\ref%
{enunderover})-(\ref{enunderover2}) (which we repeat here for
convenience,with $2\pi \tilde{J}=:J$),
\begin{eqnarray}
	S_{\text{under}} &=&\sqrt{S_{0}^{2}\left( Q\right) -4\pi ^{2}\tilde{J}^{2}}%
	=S_{0}\left( Q\right) \sqrt{1-\alpha }; \\
	S_{\text{over}} &=&\sqrt{4\pi ^{2}\tilde{J}^{2}-S_{0}^{2}\left( Q\right) }%
	=S_{0}\left( Q\right) \sqrt{\alpha -1}; \\
	\alpha &:&=\frac{4\pi ^{2}\tilde{J}^{2}}{{S_{0}^{2}}\left( Q\right) }=\alpha
	\left( Q,\tilde{J}\right) \geqslant 0,  \label{def-alpha}
\end{eqnarray}%
depending whether the BH is (stationarily) under- or over- rotating,
respectively. Conveniently, these forms of extremal Bekenstein-Hawking
entropy can be written as
\begin{equation}
	S_{\text{under(over)}}=S_{0}\sqrt{|1-\alpha |},  \label{this3}
\end{equation}%
where $\alpha >1$ resp. $\alpha <1$ for over- resp. under- rotating extremal
BHs.
By the aforementioned KN/CFT correspondence, the central charge %
\eqref{centralcharge} and the temperature \eqref{lefttemp} of the
corresponding, dual CFT can be computed to read
\begin{eqnarray}
	c &=&12\tilde{J}; \\
	T_{L} &=&\frac{1}{2\pi }\sqrt{\frac{|1-\alpha |}{\alpha }},  \label{thisss}
\end{eqnarray}%
and thus the Cardy entropy reads
\begin{equation}
	S_{CFT}=S_{0}\sqrt{|1-\alpha |}=S_{\text{under(over)}},
\end{equation}%
therefore again matching the Bekenstein-Hawking BH entropy (\ref{this3}).

\section{RFD in CFT}

\label{sec:cardy} In the previous sections, we have separately discussed the
RFD on both extremal and double-extremal black holes and the KN/CFT correspondence. As proved in \cite{Chattopadhyay:2023nfc}%
, it is here worth recalling that two asymptotically flat, extremal
(stationary) rotating BHs possessing different e.m. charges $Q$ resp. $\hat{Q%
}$, as well as different angular momenta $J$ resp. $\hat{J}:=J+\delta J$,
have the same Bekenstein-Hawking entropy when their e.m. charges are related
by RFD $\hat{Q}:=\Omega \frac{\partial S(Q,\hat{J})}{\partial Q}$. A similar duality also holds for the double-extremal black holes. As
reviewed in \cref{sec:KNBH} and \cref{kerrcftother}, the Bekenstein-Hawking
entropy of these BHs precisely matches the Cardy entropy of their dual CFTs
(according to the Kerr/CFT correspondence and its various extensions
reviewed above). This naturally leads to the question of how RFD operates on
these dual CFTs, which is the main scope of this paper. In this section, we are going to analyse that in detail for both extremal and doubly-extremal BHs.

\subsection{Central charge and Temperature}

We now consider applying the RFD to a rotating, extremal (or doubly-extremal) BH with e.m. charge
$Q$ and angular momentum $J$. From the KN/CFT correspondence, the dual CFT
possesses a temperature $T_{L}$ and central charge $c_{L}$; therefore,
the Cardy entropy is generally given by (\ref{eq:cardentropy2}), with $c_{L}$
given by (\ref{tl1}). Let us denote the central charge and temperature of
the CFT that is dual to the RFD-transformed BH as ${c}_{L}^{\prime }$ and ${T%
}_{L}^{\prime }$, with ${c}_{L}^{\prime }=12\widetilde{\hat{J},}$ where $%
\widetilde{\hat{J}}:=2\pi \hat{J}=2\pi \left( J+\delta J\right) $. For convenience, we introduce a running function $w(\alpha)$ such that
	\begin{eqnarray}\label{eq:walpha}
		w(\alpha)=  \begin{cases}
			f(\alpha) \quad \text{for an extremal BH}\\
			h(\alpha) \quad \text{for a doubly-extremal BH}\end{cases}
	\end{eqnarray}
Thus, it holds that
	\begin{equation}
		{c}_{L}^{\prime }\overset{\text{(\ref{thiss},\ref{hja})}}{=}2\pi \cdot 12\sqrt{%
			w(\alpha )}S_{0}{=}2\pi \cdot 12\sqrt{%
			\frac{w(\alpha )}{\alpha }}J\overset{\text{(\ref{tl1})}}{=}\sqrt{\frac{%
				w(\alpha )}{\alpha }}c_{L}.  \label{RFD-on-c}
	\end{equation}%
	Since the RFD preserves the Bekenstein-Hawking BH entropy, the KN/CFT
	correspondence and (\ref{eq:cardentropy2}) thus imply that
	\begin{equation}
		\frac{\pi ^{2}}{3}{c}_{L}^{\prime }{T}_{L}^{\prime }=\frac{\pi ^{2}}{3}%
		c_{L}T_{L}\Leftrightarrow {T}_{L}^{\prime }=\frac{c_{L}T_{L}}{{c}%
			_{L}^{\prime }}\overset{\text{(\ref{tl1})}}{=}\frac{JT_{L}}{\hat{J}}\overset{%
			\text{(\ref{thiss},\ref{hja})}}{=}\sqrt{\frac{\alpha }{w(\alpha )}}%
		T_{L}.  \label{RFD-on-T}
	\end{equation}%
Hence, we find that, for both extremal and doubly-extremal BHs, two different (thermal) CFTs with different temperatures
and central charges have the same Cardy entropy when their dual extremal resp. doubly-extremal
rotating BHs (through the Kerr/CFT correspondence and its generalizations)
have their e.m. charges and angular momenta related by the RFD.

\subsection{Remark}

By recalling (\ref{eq:cardentropy2}), it is clear that $S_{\text{Cardy}}^{%
	\text{KN}}$ is invariant under a simultaneous dilatation on $c_{L}$ and $%
T_{L}$ with opposite weights :%
\begin{equation}
	\left.
	\begin{array}{l}
		c_{L}\rightarrow e^{\xi }c_{L}, \\
		\\
		T_{L}\rightarrow e^{\varphi }T_{L};%
	\end{array}%
	\right\} \Rightarrow S_{\text{Cardy}}^{\text{KN}}\rightarrow S_{\text{Cardy}%
	}^{\text{KN}}\Leftrightarrow \xi =-\varphi .  \label{transf}
\end{equation}%
Therefore, it is clear that the action (\ref{RFD-on-c})-(\ref{RFD-on-T}) of
the RFD on the CFT side of the KN/CFT correspondence, namely on $c$ and $%
T_{(L)}$ of the dual CFT, is a particular case of the dilatational symmetry (%
\ref{transf}) of $S_{\text{Cardy}}^{\text{KN}}$, with%
	\begin{equation}
		\xi =-\varphi =\frac{1}{2}\ln \left( \frac{w\left( \alpha \right) }{\alpha }%
		\right) .
	\end{equation}%
	By definition (cf. (\ref{eq:walpha})), $w(\alpha)=f(\alpha)$ defined in (\ref{eq:falpha}) for extremal, and $w(\alpha)=h(\alpha)$ as in (\ref{eq:ha}) for doubly-extremal BH.

\subsection{(Asymptotic) Density of states}

We further delve into the action of the RFD on the density of states of the
thermal CFTs which are dual, through the KN/CFT correspondence, to two
RFD-dual rotating extremal (or doubly-extremal) BHs. With the help of the thermodynamic relation %
\cref{eq:thermo} between temperature $T$ and the energy $\Delta $, we can
deduce the expression of the RFD-transformed energy,
	\begin{equation}
		\Delta ^{\prime }=\sqrt{\frac{\alpha }{w(\alpha )}}\Delta .
		\label{eq:deltarel}
	\end{equation}%
The holomorphic part of the RFD-transformed density of states can be written
as\footnote{%
	In the present paper, we will denote the imaginary unit by $\mathbf{i}$.}
\begin{equation}
	\rho ^{\prime }(\Delta ^{\prime })=\int e^{2\pi \mathbf{i}g^{\prime }\left(
		\tau \right) }Z(-1/\tau )d\tau ,  \label{eq:rho3}
\end{equation}%
where
\begin{equation}
	g^{\prime }(\tau ):=-\tau \Delta ^{\prime }+{\frac{c^{\prime }}{24\tau }}+{%
		\frac{c^{\prime }\tau }{24}}.  \label{eq:gtp}
\end{equation}%
As in \cite{Cardy:1986ie,cardy2}, the indefinite integral appearing in (\ref%
{eq:rho3}) can be evaluated by using the steepest descent method; for a
large value of $\Delta ^{\prime }$, the exponent is extremised at the point
	\begin{equation}
		\tau _{0}^{\prime }=\sqrt{\frac{c^{\prime }}{-24\Delta ^{\prime }+c^{\prime }%
		}}\approx \mathbf{i}\sqrt{\frac{w(\alpha )}{\alpha }}\sqrt{\frac{c}{24\Delta
		}},
\end{equation}%
to be contrasted with $\tau _{0}\approx \mathbf{i}\sqrt{\frac{c}{24\Delta }}$%
. Therefore, one obtains the relation
	\begin{equation}
		\tau _{0}^{\prime }=\tau _{0}\sqrt{\frac{w(\alpha )}{\alpha }}.
		\label{tthis}
\end{equation}%
With all this together, \cref{eq:gtp} gets simplified as follows :%
\begin{equation}
	g^{\prime }(\tau )\approx g^{\prime }(\tau _{0}^{\prime })=-\tau
	_{0}^{\prime }\Delta ^{\prime }+{\frac{c^{\prime }}{24\tau _{0}^{\prime }}}+{%
		\frac{c^{\prime }\tau _{0}^{\prime }}{24}}\overset{\text{(\ref{eq:deltarel}%
			),(\ref{tthis})}}{=}-\tau _{0}\Delta +\frac{c}{24\tau _{0}}+\frac{w(\alpha )%
		}{\alpha }{\frac{c\tau _{0}}{24}}.
\end{equation}%
Therefore, the only modification at the saddle point value of the exponent
comes through the last term.

Finally, we can write the density of states of the thermal CFT which is dual
to the RFD-transformed extremal (or doubly-extremal) rotating BH, for large\footnote{%
	Large value of $\Delta ^{\prime }$ $\implies $ large value of $\Delta $ as
	they are related through \cref{eq:deltarel}, since the ratio $\alpha
		/w(\alpha )$ is always finite and non-negative.} $\Delta ^{\prime }$, as
	\begin{eqnarray}
		\rho ^{\prime }(\Delta ^{\prime }) &\approx &\exp \left( 4\pi \sqrt{\frac{%
				c\Delta }{24}}\right) \exp \left( -{\frac{2\pi c}{24}}\sqrt{\frac{c}{%
				24\Delta }}{\frac{w(\alpha )}{\alpha }}\right) Z\left( -1/(\mathbf{i}\sqrt{%
			w(\alpha )/\alpha }\sqrt{c/24\Delta })\right)  \notag \\
		&\approx &\exp \left( 2\pi \sqrt{\frac{c\Delta }{6}}\right) =\rho (\Delta ).
		\label{rho}
\end{eqnarray}%
Thus, one can conclude that \textit{the asymptotic density of states remains
	invariant under RFD}, which is expected due to the invariance of the Cardy
entropy itself. \ Note that, in order to derive (\ref{rho}), the following
approximations have been used\footnote{%
	In (\ref{appr2}) the modular symmetry of $Z$ has been used (see the \textit{r%
		\'{e}sum\'{e}} in App. \ref{app:kerrcardy} and Refs. therein).} :%
\begin{eqnarray}
		&&\exp \left( -{\frac{2\pi c}{24}}\sqrt{\frac{c}{24\Delta }}{\frac{w(\alpha )%
			}{\alpha }}\right) \overset{\Delta \gg 1}{\approx }1;  \label{appr1} \\
		&&Z\left( -1/(\mathbf{i}\sqrt{w(\alpha )/\alpha }\sqrt{c/24\Delta })\right)
		\overset{\Delta \gg 1}{\approx }1.  \label{appr2}
\end{eqnarray}%
This is exactly the line of thought that one uses in order to derive the
Cardy formula for CFT; see \cref{app:kerrcardy}, in which the treatment
of \cite{Cardy:1986ie,cardy2} is sketchily reviewed.

\section{Discussion and Outlook}

\label{sec:conclusion}

Rotational Freudenthal duality (RFD) is a generalization of Freudenthal
duality for asymptotically flat, rotating, stationary, generally dyonic
extremal BHs, introduced in \cite{Chattopadhyay:2023nfc}. It yields that two
(under- or over-) rotating extremal BHs have the same Bekenstein-Hawking
entropy in spite of having their e.m. charges related by a non-linear
relation when they have certain distinct (uniquely and analytically
determined) angular momenta. In this paper, we begin by completing the analysis done in \cite{Chattopadhyay:2023nfc}. In \cref{app:anotherrfd} we have proved that there exists a unique,
	Freudenthal dual BH for \textit{doubly-extremal} rotating BHs, and we have
	also found a `spurious' (but `golden'!) branch of solutions in the
	non-rotating limit. Intriguingly, we have also obtained that the RFD will
	\textit{necessarily} boost a rotating doubly-extremal BH (with any ratio of $%
	S_{0}\left( Q\right) $ and $J$) to an over-rotating regime. All these
		results surely warrant some further future investigation.

In the subsequent part of the paper, we have primarily aimed at understanding
the action of the RFD on the microscopic entropy of extremal (or doubly-extremal) Kerr-Newman
(KN) BHs. Using the Kerr/CFT correspondence (and its further extensions to
Maxwell and scalar fields, thus applicable to Maxwell-Einstein-scalar
systems, as the ones occurring in the purely bosonic sector of
four-dimensional supergravities), we have investigated how the RFD
influences the density of states of a thermal CFT, which is dual to the
extremal (or doubly-extremal) rotating BH under consideration.

On the CFT side of the KN/CFT correspondence, we found that the RFD relates
two two-dimensional thermal CFTs, whose density of states is the same,
though they have different temperatures and central charges (but the same
Cardy entropy!). More explicitly, by recalling (\ref{RFD-on-c}) and (\ref%
{RFD-on-T}), we have determined that the action of the RFD on the (left)
central charge and on the (left) temperature reads as follows :%
\begin{eqnarray}
		c_{L} &\rightarrow &c_{L}^{\prime }:=\sqrt{\frac{w\left( \alpha \right) }{%
				\alpha }}c_{L}; \\
		T_{L} &\rightarrow &T_{L}^{\prime }:=\sqrt{\frac{\alpha }{w\left( \alpha
				\right) }}T_{L}\overset{\text{(\ref{thisss})}}{=}\frac{1}{2\pi }\sqrt{\frac{%
				\alpha }{w\left( \alpha \right) }}\sqrt{\frac{\left\vert 1-\alpha
				\right\vert }{\alpha }}=\frac{1}{2\pi }\sqrt{\frac{\left\vert 1-\alpha
				\right\vert }{w\left( \alpha \right) }}.
\end{eqnarray}%
It is amusing to observe that the net effect of the RFD on the temperature
is the replacement of $\alpha $ with $w(\alpha)$(=$f(\alpha)$ defined in (\ref{eq:falpha}) or $h(\alpha)$ defined in (\ref{eq:ha})) in the denominator of $T_{L}$ (\ref{thisss}) :%
	\begin{equation}
		T_{L}=\frac{1}{2\pi }\sqrt{\frac{\left\vert 1-\alpha \right\vert }{\alpha }}%
		\rightarrow T_{L}^{\prime }=\frac{1}{2\pi }\sqrt{\frac{\left\vert 1-\alpha
				\right\vert }{w\left( \alpha \right) }}.
\end{equation}
One should note two interesting domains of $\alpha$.

\begin{wrapfigure}[13]{l}{0.55\textwidth}
	\includegraphics[width=\linewidth]{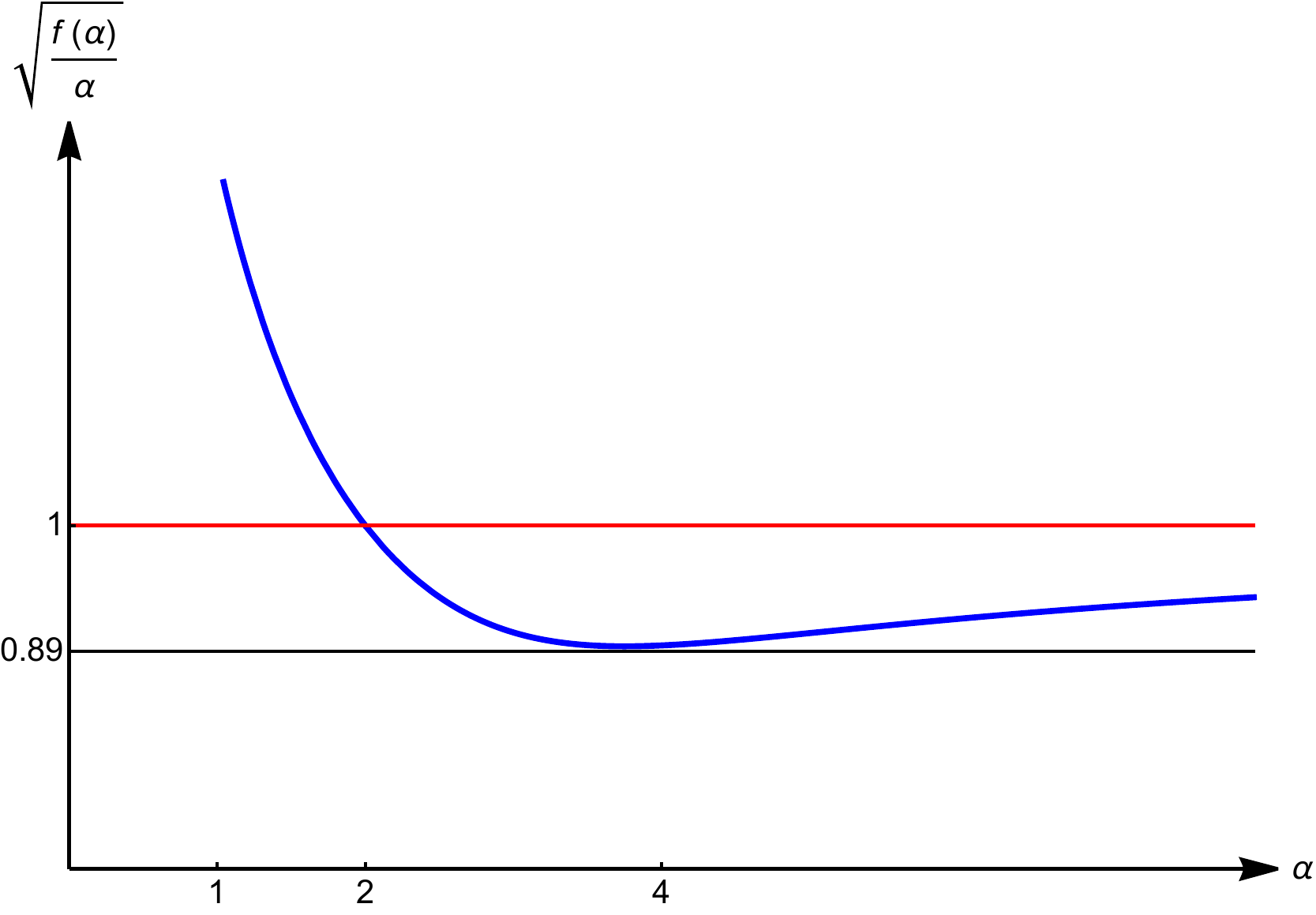}
	\caption{Plot of $\sqrt{\frac{f(\alpha )}{\alpha }}$ shown in blue with a minima at $\alpha\approx 3.75$.}
	\label{fig:fabya}
\end{wrapfigure}

\begin{enumerate}
	\item \textbf{Extremal Black hole:} The RFD-self-dual value $\alpha =2$, at which the RFD
	trivializes to the identity (because $w(2)\equiv f(2)=2$; cf. (\ref{plott})) : it
	corresponds to a peculiar over-rotating extremal BH, with $%
	J^{2}=2S_{0}^{2}(Q)$, discussed in Sec. 2.2 of \cite{Chattopadhyay:2023nfc}.\\
	\textbf{Doubly-extremal Black hole:} On the contrary to its extremal counterpart, nothing similar happens for $w(\alpha)= h(\alpha)$, which nicely ties in to the fact that \emph{RFD transformed doubly-extremal KN BH is always over-rotating} cf. section \ref{sec:no_or_KNBH}.
	
	\item \textbf{Extremal Black hole:} The second one is the interval $\alpha \in \lbrack 0,2)$, in which $%
	\frac{f\left( \alpha \right) }{\alpha }<1$. Therefore, one could have na%
	\"{\i}vely made the observation that for this range of $\alpha $ the RFD
	might map a unitary CFT (with $c>1$) to a non-unitary one (with $c<1$) by
	virtue of the formula (\ref{RFD-on-c}); this would have resulted in a
	conceptual issue, since one of the core assumptions of the Kerr/CFT
	correspondence (and its extensions) is that the dual CFT is unitary. But a
	careful look at the values of $\sqrt{\frac{f\left( \alpha \right) }{\alpha }}
	$ reveals that this function has a (global) minimum at $\alpha \approx 3.75$%
	, at which $\sqrt{\frac{f\left( \alpha \right) }{\alpha }}\approx 0.894$, as
	shown in figure \ref{fig:fabya}. Since the Kerr/CFT correspondence works in
	the large $c$ limit (i.e., for $c\gg 1$), the central charge will still
	remain large under the action of the RFD. This observation raises the
	interesting question whether one could analytically extend the definition of
	the RFD for all values of $c$ of a CFT and analyze the effect of RFD itself
	on the unitarity of the theory. We leave this to further future
	investigation.\\
	\textbf{Double-extremal Black hole:} In this case ${h(\alpha)\over \alpha}>1\,\,\forall\, \alpha\in (0,\infty)$ implying no such unitarity issue as above. This can easily be observed from (\ref{eq:hrel}) and  the aforementioned observation that $h(\alpha)>0\,\forall\,\alpha\in(0,\infty)$, which implies $(1+h(\alpha))^2>1$. Therefore $\forall\,\alpha\in(0,\infty)$ we have 
		\begin{eqnarray}
			0<1+\alpha-h(\alpha)<1\implies\alpha-h(\alpha)<0\implies{h(\alpha)\over \alpha}>1.
	\end{eqnarray}
\end{enumerate}

Though our analysis did not address the effect of the RFD on the degeneracy
of BH microstates, still it employed the KN/CFT correspondence. Explicitly, we exploit the fact that the microscopic/statistical (Cardy) entropy of a two-dimensional
thermal CFT (on the CFT side of the correspondence) matches the
macroscopic/thermodynamical (Bekenstein - Hawking) entropy of the
asymptotically flat KN extremal (or doubly-extremal) BH under consideration (on the BH side of
the correspondence). An intriguing further development may concern the study
of the action of the RFD on the bound states of $D$-branes \cite%
{Strominger:1996sh} or $M$-branes \cite{Juan-Maldacena-1997} that account
for the Bekenstein-Hawking BH entropy itself.

It will also be interesting to extend the present investigation to
asymptotically AdS, KN extremal (or doubly-extremal) BH solutions, in which one would be able to
apply the AdS/CFT correspondence and the KN/CFT correspondence to the
(spacial) asymptotical and near-horizon limits of the scalar flow,
respectively. This research venue appears particularly promising, and we
keep it as a future endeavour.

\section*{\noindent \textbf{Acknowledgments}}

The authors would like to thank the anonymous referee for constructive criticism, which helped to improve the presentation of this paper. AC would like to thank Prof. M. Khalkhali for illuminating discussions.

The work of AC is supported by the European Union's Horizon 2020 research
and innovation programme under the Marie Sk\l {}odowska Curie grant
agreement number 101034383. The work of TM is supported by
the grant RJF/2022/000130 from the Science and Engineering Research Board (SERB), India. The work of AM is supported by a
\textquotedblleft Maria Zambrano\textquotedblright\ distinguished researcher
fellowship at the University of Murcia, Spain, financed by the European
Union within the NextGenerationEU programme. This article is based upon work from COST Action CaLISTA CA21109 supported by COST (European Cooperation in Science and Technology).

\appendix

\section{Cardy formula : a quick \textit{r\'{e}sum\'{e}}}

\label{app:kerrcardy}

The Cardy formula plays a crucial role in our analysis. In this appendix, we
provide a concise overview of the derivation of the Cardy formula \cite%
{Cardy:1986ie,cardy2}, which is employed to calculate the density of states
in a two-dimensional CFT and subsequently determine its statistical entropy.

Thus, we start with a two-dimensional CFT endowed with the usual pair of
Virasoro algebras governing the left- and right- moving modes, each
characterized by a central charge denoted as $c$. Exploiting the conformal
symmetry, it becomes possible to map the theory from a plane to a cylinder
without loss of generality. Moreover, one can compactify the cylinder into a
torus by using a complex modulus $\tau =\tau _{1}+\mathbf{i}\tau _{2}$. The
torus partition function can then be written as follows
\begin{equation}
	Z(\tau ,\bar{\tau})=\text{Tr}\,e^{2\pi \mathbf{i}\tau L_{0}}e^{-2\pi \mathbf{%
			i}\bar{\tau}\bar{L}_{0}},
\end{equation}%
with $L_{0}$ and $\bar{L}_{0}$ being the global generators of the
holomorphic and the anti-holomorphic part. Since the trace is over all
states in the Hilbert space, one can further write the partition function as
\begin{equation}
	Z(\tau ,\bar{\tau})=\text{Tr}\,e^{2\pi \mathbf{i}\tau L_{0}}e^{-2\pi \mathbf{%
			i}\bar{\tau}\bar{L}_{0}}=\sum_{\{\Delta ,\bar{\Delta}\}}\rho (\Delta ,\bar{%
		\Delta})e^{2\pi \mathbf{i}\Delta \tau }e^{-2\pi \mathbf{i}\bar{\Delta}\bar{%
			\tau}},
\end{equation}%
with $\rho (\Delta ,\bar{\Delta})$ being the density of states and $\Delta $
denoting the conformal scaling dimension. Using the \textit{residue trick},
one can then express the density of states in terms of the partition
function as
\begin{equation}
	\rho (\Delta ,\bar{\Delta})={\frac{1}{(2\pi \mathbf{i})^{2}}}\int {\frac{dq}{%
			q^{\Delta +1}}}{\frac{d\bar{q}}{\bar{q}^{\bar{\Delta}+1}}}Z(q,\bar{q}),
	\label{eq:rho1}
\end{equation}%
where $q=e^{2\pi \mathbf{i}\tau }$ and $\bar{q}=e^{2\pi \mathbf{i}\bar{\tau}%
} $ and the contour encloses the point $(q,\bar{q})\equiv (0,0)$.

A crucial step in deriving the Cardy formula is the definition of the
function
\begin{equation}
	Z_{0}(\tau ,\bar{\tau}):=\text{Tr}\,e^{2\pi \mathbf{i}(L_{o}-{\frac{c}{24}}%
		)\tau }e^{-2\pi \mathbf{i}(\bar{L}_{o}-{\frac{c}{24}})\bar{\tau}},
\end{equation}%
which enjoys modular invariance under the transformation $\tau \rightarrow -{%
	1/\tau }$. Therefore, one can recast the density of states given by %
\cref{eq:rho1} as
\begin{equation}
	\rho (\Delta ,\bar{\Delta})={\frac{1}{(2\pi \mathbf{i})^{2}}}\int {\frac{dq}{%
			q^{\Delta +1}}}{\frac{d\bar{q}}{\bar{q}^{\bar{\Delta}+1}}}Z_{0}(q,\bar{q})q^{%
		\frac{c}{24}}\bar{q}^{-{\frac{c}{24}}}.
\end{equation}%
From now onwards, we will focus on the holomorphic parts only assuming a
similar treatment with the anti-holomorphic sector. Using the modular
symmetry of $Z_{0}(\tau )$, one obtains
\begin{equation}
	Z_{0}(\tau )=Z_{0}\left( -{\frac{1}{\tau }}\right) =e^{{\frac{2\pi \mathbf{i}%
				c}{24\tau }}}Z\left( -{\frac{1}{\tau }}\right) .
\end{equation}%
Using all the above relations together, we have the holomorphic part of the
density of states rewritten as
\begin{equation}
	\rho (\Delta )=\int e^{2\pi \mathbf{i}g(\tau )}Z(-1/\tau )d\tau ,
	\label{eq:rho2}
\end{equation}%
with the definition
\begin{equation}
	g(\tau ):=-\tau \Delta +{\frac{c}{24\tau }}+{\frac{{c\tau }}{24}}.
\end{equation}%
The whole point of using the modular symmetry is to claim the fact that $%
Z(-1/\tau )$ approaches a constant value for large values of $\tau _{2}$,
which is evident by the construction. Therefore, we can now use the \textit{%
	steepest descent method} to calculate the integral in \cref{eq:rho2}. In
order to proceed with the steepest descent, one has to look for the value of
$\tau =\tau _{0}$ such that the power of the exponent in \cref{eq:rho2} is
extremised. This implies that for $dg/d\tau \rightarrow 0$ one obtains
\begin{equation}
	\tau _{0}=\sqrt{\frac{c}{-24\Delta +c}}\approx \mathbf{i}\sqrt{\frac{c}{%
			24\Delta }},
\end{equation}%
where we have assumed $\Delta $ to be large ($\Delta \gg c$) or, in other
words, we are looking for asymptotic density of states through \cref{eq:rho2}%
. Therefore, at the saddle point for large $\Delta $ values, we have
\begin{eqnarray}
	\rho (\Delta ) &\approx &\exp \left( 4\pi \sqrt{\frac{c\Delta }{24}}\right)
	\exp \left( -{\frac{2\pi c}{24}}\sqrt{\frac{c}{24\Delta }}\right) Z\left( -%
	\frac{1}{i\sqrt{c/24\Delta }}\right)  \notag \\
	&\approx &\exp \left( 2\pi \sqrt{\frac{c\Delta }{6}}\right) Z(i\infty
	)\approx \exp \left( 2\pi \sqrt{\frac{c\Delta }{6}}\right) ,
\end{eqnarray}%
where%
\begin{equation}
	\exp \left( -{\frac{2\pi c}{24}}\sqrt{\frac{c}{24\Delta }}\right) \overset{%
		\Delta \gg 1}{\approx }1,
\end{equation}%
and modular symmetry implies that%
\begin{equation}
	Z(i\infty )\overset{\Delta \gg 1}{\approx }1.
\end{equation}

Thus, without even knowing many details of the theory, one can arrive at the
definition of entropy as
\begin{equation}
	S_{\text{Cardy}}=\log \left( \rho (\Delta )\right) =2\pi \sqrt{\frac{c\Delta
		}{6}}.
\end{equation}%
Since in this formula $c$ is the central charge and $\Delta $ is the
analogue of energy, the temperature of the CFT can be defined through the
relation
\begin{equation}
	{\frac{1}{T}}={\frac{dS_{\text{Cardy}}}{d\Delta }}=\pi \sqrt{\frac{c}{%
			6\Delta }}.  \label{eq:thermo}
\end{equation}%
This brings us to the famous Cardy formula
\begin{equation}
	S_{\text{Cardy}}={\frac{\pi ^{2}}{3}}cT.
\end{equation}

\bibliographystyle{jhep}
\bibliography{cardy_F}
{}

\end{document}